**A kinetic model predicts that tau-catalyzed Aβ nucleation can raise oligomer burden without increasing plaque volume**

Andrey V. Kuznetsov

Department of Mechanical and Aerospace Engineering, North Carolina State University,

Raleigh, NC 27695-7910, USA; email: avkuznet@ncsu.edu

## Abstract

Tau and amyloid-β (Aβ) deposits co-occur in Alzheimer's disease, and tau fibrils have recently been shown *in vitro* to catalyze Aβ42 primary nucleation in a fold-specific manner. Whether tau-catalyzed Aβ nucleation is relevant *in vivo* remains unknown. A two-compartment kinetic model is developed here, describing tau aggregation within the neuronal soma and Aβ aggregation in the surrounding interstitium, coupled by a heterotypic nucleation term calibrated against the published *in vitro* measurements. The adjustable coupling parameter is the fraction $f_{acc}$ of somatic tau fibrillar material accessible to interstitial Aβ. Homotypic Aβ secondary nucleation is restricted to the accessible surface of the consolidating plaque rather than scaling with total fibril mass, which preserves the *in vitro* calibration of the rate constant while preventing the surface term from growing without bound over decades. Three model predictions emerge. First, the model predicts that tau-catalyzed nucleation can increase the soluble Aβ oligomer concentration and the number of Aβ fibrillar species, by factors of 4.6 and 2.0, respectively, at complete accessibility, while leaving total Aβ fibril mass and plaque volume unchanged, since in the absence of clearance these are fixed by the monomer supply; oligomer burden and visible plaque burden are therefore decoupled. Second, the predicted effect requires substantial accessibility: a 50 % increase in accumulated oligomer exposure needs $f_{acc} \approx 0.035$–$0.15$, which is difficult to reconcile with tangles enclosed by an intact neuronal membrane and points instead to neuronal lysis and the formation of extracellular ghost tangles. Third, within the model, physiologically relevant oligomer dissociation and proteolytic turnover each suppress accumulated exposure, by approximately three orders of magnitude and, at zero tau accessibility, about seventyfold, respectively, and remove the effect on a biological-age measure entirely. A within-dataset closed-system consistency check reproduces PHF-tau lag-time acceleration within 6–16 % in the calibration dataset, although the absolute aggregation timescale is not reproduced. These quantitative results should be interpreted as model predictions rather than validated *in vivo* estimates.

**Abbreviations**

| | |
|---|---|
| AD | Alzheimer’s disease |
| Aβ | amyloid beta |
| CV | control volume |
| ECS | extracellular space |
| PHF | paired helical filament |
| CTE | chronic traumatic encephalopathy |

## 1. Introduction

Alzheimer's disease (AD) is characterized pathologically by the accumulation of amyloid-β (Aβ) in extracellular plaques and tau in intracellular neurofibrillary tangles. These two proteinopathies have traditionally been considered as spatially distinct processes, with Aβ pathology often placed upstream of tau pathology. Increasing experimental evidence, however, indicates that Aβ and tau are not independent components of the disease process but can interact directly at locations where their pathological distributions overlap. Aβ and tau co-localize within diseased synaptic terminals (Fein et al., 2008), and soluble forms of the two proteins can form heterocomplexes that promote their conversion into insoluble aggregates (Guo et al., 2006). Functional studies provide further evidence for such coupling: depletion of tau in human neurons markedly attenuates Aβ-induced neuronal dysfunction and neurodegeneration (Ng et al., 2024), while structural analyses suggest that Aβ and tau may share complementary interaction surfaces capable of facilitating heterotypic association (Griner et al., 2019). Together, these observations suggest that direct molecular communication between Aβ and tau may contribute importantly to AD pathology.

Recent work by (Mosconi et al., 2026) provides a specific kinetic mechanism for this interaction. Using disease-associated tau fibrils, they demonstrated that preformed tau aggregates can directly catalyze the primary nucleation of Aβ42. This heterotypic nucleation process is distinct from the fibril-surface secondary nucleation that drives autocatalytic Aβ self-replication (Cohen et al., 2013). The catalytic activity depended strongly on tau fibril structure: chronic traumatic encephalopathy (CTE)-associated tau fibrils, which possess a fold distinct from the paired helical filament (PHF) structure characteristic of AD (Fitzpatrick et al., 2017), (Falcon et al., 2019), accelerated Aβ42 nucleation more strongly than PHF fibrils (Mosconi et al., 2026). The experiments further indicated enzyme-like behavior, including saturation and positive cooperativity, and identified the exposed tau region spanning approximately residues 351–378 as a candidate interaction site. Importantly, tau cross-seeding also produced shorter Aβ42 fibrils, consistent with enhanced nucleation generating a larger population of growing fibrils and thereby reducing the amount of free Aβ monomer available for elongation. These findings suggest that tau may influence not only the rate of Aβ aggregation but also the distribution of Aβ among monomeric, oligomeric, and fibrillar states.

This distinction is potentially important because the biological consequences of aggregation are not determined solely by total fibrillar mass. Soluble oligomeric forms of both Aβ and tau are strongly implicated in synaptic dysfunction and neuronal injury, whereas plaque and tangle burden provide only an incomplete measure of the underlying toxic protein population. Thus, an increase in nucleation could be pathologically important even if it caused little change in total fibril mass, provided that it increased the number of aggregates, altered their characteristic size, or changed the concentration and persistence of soluble oligomeric intermediates. Protein turnover adds another level of competition to this process: oligomer dissociation and proteolytic degradation can remove potentially toxic species, whereas impairment of protein-clearance mechanisms can permit their accumulation over the decades-long course of AD. A quantitative description of Aβ–tau coupling therefore requires simultaneous consideration of nucleation, oligomer formation and dissociation, fibril growth, fibril mass, and protein clearance.

Despite the growing experimental evidence for direct Aβ–tau interactions, their consequences for long-term coupled aggregation dynamics remain incompletely quantified. In particular, it is not evident whether tau-catalyzed Aβ nucleation should primarily increase total plaque burden or instead redistribute a given Aβ mass among a larger number of shorter fibrils while altering the transient oligomer population. Nor is it clear how strongly these effects depend on the fraction of intracellular tau fibrillar material that becomes accessible to extracellular or synaptic Aβ, or how oligomer dissociation and protein degradation modify the resulting cumulative neurotoxic burden.

To address these questions, the present study develops a coupled two-compartment mathematical model of tau and Aβ aggregation. The framework extends previously published Aβ plaque-growth and combined Aβ-tau neurotoxicity formulations (Kuznetsov, 2024), (Kuznetsov, 2025a), (Kuznetsov, 2026a); the new contribution is coupling the aggregation modules through fold-dependent heterotypic nucleation and restricting homotypic Aβ secondary nucleation to accessible plaque surface. Tau aggregation is described within the intracellular volume of a representative neuronal soma, whereas Aβ aggregation is modeled within the surrounding extracellular/interstitial tissue region. The two compartments are coupled through an effective concentration of tau fibrillar material that becomes accessible to Aβ at sites of pathological overlap. Aβ aggregation includes intrinsic primary nucleation, Aβ-fibril-surface secondary nucleation, oligomer dissociation, oligomer-to-fibril conversion, fibril elongation, fragmentation, and protein turnover. Tau-mediated Aβ aggregation is introduced as an additional heterotypic primary-nucleation pathway, with its fold-dependent coefficient calibrated directly from the seed-concentration-dependent kinetic measurements of (Mosconi et al., 2026), rather than through an independently assumed catalytic rate constant.

The model is used to determine how tau fibril accessibility modifies free Aβ monomer concentration, soluble Aβ oligomer concentration, fibril number, fibril mass, plaque burden, and accumulated oligomer exposure. Particular attention is given to whether tau-mediated nucleation redistributes Aβ toward a larger number of shorter fibrils, as suggested experimentally by (Mosconi et al., 2026), and whether this redistribution can substantially increase oligomer-mediated neurotoxicity without a proportional increase in plaque mass. The framework further incorporates the accumulated exposure to both Aβ and tau oligomers as a model-derived measure of cumulative neurotoxicity and examines how oligomer dissociation and protein degradation alter this burden. In this way, the model provides a quantitative framework for investigating how direct tau–Aβ coupling, tau fibril structure, and the integrity of protein-clearance mechanisms may jointly influence the long-term molecular progression of AD.

## 2. Materials and models

### 2.1. Mathematical framework for tau aggregation

Mass conservation for tau monomers, oligomers, and fibrillar mass, together with a balance for fibril number, is formulated within the intracellular volume of a representative neuronal soma. The soma therefore constitutes the tau control volume, $V_{tau} = V_{soma}$. Using the value already adopted in Table 2, $V_{soma} =$ 4.19×10³ μm³, the representative soma has a diameter of approximately 20 μm when approximated as a sphere. All tau species are assumed to be spatially homogeneous within this lumped intracellular compartment. Tau monomers are supplied to the soma at the rate $q_{tau}$, so that $q_{tau} / V_{soma}$, after application of the conversion factor $a_{21}$, represents the corresponding volumetric source term. Applying conservation of free tau monomers yields the following governing equation:

$$\frac{dm_{tau}}{dt} = -k_{o1,tau} m_{tau}(t) - k_{o2,tau} m_{tau}^2(t) M_{tau}(t) + k_{d1,tau} S_{tau}(t) + k_{d2,tau} S_{tau}(t) M_{tau}(t) - 2k_{+,tau} m_{tau}(t) P_{tau}(t)$$

$$-k_{conv,tau} m_{tau}(t) S_{tau}(t) - m_{tau}(t) \frac{\ln(2)}{T_{1/2,m,tau}} + \frac{q_{tau}}{V_{soma}} a_{21} \qquad (1)$$

where $t$ is the time elapsed since AD onset. The first term on the right-hand side of Eq. (1) represents tau monomer consumption by effective first-order primary nucleation, and the second describes fibril-surface-catalyzed secondary nucleation of tau monomers into on-pathway oligomers. The third and fourth terms return monomers through spontaneous and fibril-surface-catalyzed oligomer dissociation, respectively. The fifth term describes monomer consumption by fibril elongation at the two fibril ends, and the sixth represents conversion of an oligomer into an elongation-competent fibril through monomer addition. The seventh term accounts for proteolytic clearance of tau monomers. Finally, the eighth term represents tau monomer supply to the soma at the rate $q_{tau}$; division by $V_{soma}$ and multiplication by $a_{21}$ convert this supply to a volumetric concentration source.

In the present coarse-grained model, an oligomer is interpreted as an activated monomer and is assumed to carry the same mass as a free monomer (Morris et al., 2008), (Iashchishyn et al., 2017). This represents a considerable simplification of the oligomer formation mechanism. The powers of monomer concentration appearing in the nucleation rate laws are interpreted as effective kinetic reaction orders rather than literal oligomer stoichiometries. Applying the conservation principle to the number of free on-pathway tau oligomers within the CV and normalizing the resulting equation by the CV's volume yields:

$$\frac{dS_{tau}}{dt} = k_{o1,tau} m_{tau}(t) + k_{o2,tau} m_{tau}^2(t) M_{tau}(t) - k_{d1,tau} S_{tau}(t) - k_{d2,tau} S_{tau}(t) M_{tau}(t)$$

$$-k_{conv,tau} m_{tau}(t) S_{tau}(t) - S_{tau}(t) \frac{\ln(2)}{T_{1/2,S,tau}} \qquad (2)$$

The first two terms on the right-hand side of Eq. (2) represent the formation of tau oligomers through effective first-order primary nucleation and fibril-surface-catalyzed secondary nucleation, respectively. These terms are equal in magnitude and opposite in sign to the corresponding first two terms on the right-hand side of Eq. (1). The third and fourth terms describe oligomer loss due to spontaneous dissociation into monomers and fibril-catalyzed dissociation, respectively, and are equal in magnitude but opposite in sign to the corresponding terms in Eq. (1). The fifth term describes the loss of oligomers through their conversion into elongation-competent fibrils. The sixth term accounts for oligomer loss due to proteolytic degradation, the rate of which is governed by the oligomer half-life, $T_{1/2,S,tau}$.

Applying the conservation principle to the number of tau fibrillar species of all lengths and normalizing by the soma volume yields the governing equation for the total tau fibril concentration, regardless of length:

$$\frac{dP_{tau}}{dt} = k_{conv,tau} m_{tau}(t) S_{tau}(t) + k_{-,tau} \max\left[M_{tau}(t) - n_{min,tau} P_{tau}(t), 0\right] \qquad (3)$$

The first term on the right-hand side of Eq. (3) corresponds to the sixth term in Eq. (1) and the fifth term in Eq. (2), but with opposite signs, and represents the conversion of oligomers into elongation-competent fibrils. This process is assumed to proceed via monomer addition, whereby one monomer and one oligomer combine to produce a new fibril. Fibril degradation is neglected in this equation because fibrillar tau is assumed to be cleared much more slowly than monomeric tau (Majumdar et al., 2008). The second term on the right-hand side of Eq. (3) represents the generation of new fibrils through fragmentation of existing ones. (Kundel et al., 2018) provide direct evidence that full-length tau fibrils undergo slow spontaneous fragmentation, whereas the isotope-labeling experiments of (Rodriguez Camargo et al., 2021) indicate that soluble oligomers are generated predominantly from monomers by fibril-surface secondary nucleation. Fragmentation is therefore retained as a source of new fibril particles in Eq. (3), but not as a direct source of the soluble oligomer pool in Eq. (2). Fragmentation is represented with a minimum stable fibril size, so that the fragmentation term in Eq. (3) is written as $k_{-,tau}$ max ($M_{tau} - n_{min,tau} P_{tau}$, 0) rather than $k_{-,tau} M_{tau}$, where $n_{min,tau}$ is the smallest number of tau monomers that a fragment must contain to remain elongation-competent. Without this constraint, fragmentation proceeds without a lower bound: because $P_{tau}$ grows as the time integral of $M_{tau}$ while $M_{tau}$ grows linearly, the mean fibril size approaches $M_{tau}/P_{tau} = 2/(k_{-,tau}\, t)$, which after 33 years is approximately 34 monomers, or about 10 nm. That is shorter than the smallest aggregate resolvable in the imaging experiments from which $k_{-,tau}$ was obtained (100 nm, corresponding to 340 monomers), and far shorter than the 900–1000 nm filaments measured there. The unbounded form therefore extrapolates an *in vitro* fragmentation rate roughly 180-fold beyond its measurement window. Setting $n_{min,tau} = 2\times10^3$ monomers (approximately 590 nm) as the effective fragmentation cutoff, consistent with the neurofibrillary-tangle morphology used to calibrate $q_{tau}$ in Eq. (50), gives $M_{tau}/P_{tau} \approx 754$ monomers (approximately 222 nm) after 33 years. This mean can fall below the fragmentation cutoff because oligomer-to-fibril conversion can still generate new fibrils; the cutoff does not impose a minimum size on every fibril. The predictions are insensitive to the specific value: for any $n_{min,tau}$ greater than about $10^3$, fragmentation ceases to contribute to $P_{tau}$ and the fibril number is set entirely by oligomer-to-fibril conversion, so that $P_{tau}$, $M_{tau}/P_{tau}$ and $m_{tau}$ are unchanged.

The mass of fibrils within the CV is characterized by the total number of tau monomers incorporated into them. Applying conservation of this quantity and normalizing by the CV's volume yields the following equation:

$$\frac{dM_{tau}}{dt} = 2k_{+,tau} m_{tau}(t) P_{tau}(t) + 2k_{conv,tau} m_{tau}(t) S_{tau}(t) \tag{4}$$

In Eq. (4), the first term on the right-hand side describes fibril elongation through monomer addition, which can occur at either tip of the fibril. The second term describes the conversion of oligomers into fibrils, a process that involves monomer addition to the oligomer. The factor of 2 in this term reflects the incorporation of both a monomer and an oligomer into the fibril mass during this event.

Because the pathophysiological processes of AD unfold over decades (Mark and Brehmer, 2022), the long-term behavior of tau concentrations in the CV is effectively independent of initial conditions on this timescale, and all tau species can therefore be set to zero at the time of AD onset, $t = 0$. This leads to the following initial conditions for Eqs. (1)-(4):

At $t=0$: $m_{tau}=0$, $S_{tau}=0$, $P_{tau}=0$, $M_{tau}=0$ (5)

### 2.2. Mathematical framework for Aβ aggregation and tau-mediated coupling

Aβ aggregation is modeled in a second control volume representing the local extracellular/interstitial tissue region surrounding the representative neuron. To preserve the mesoscopic length scale used in the previous formulation, the effective Aβ control volume is taken as $V_{A\beta}=L^3$, where $L$ = 50 μm is approximately half the mean inter-plaque spacing. The soma occupies only a small fraction of this mesoscopic region and is treated separately as the intracellular tau compartment. Aβ monomer production at neuronal membranes is retained in the existing flux-equivalent form $q_{A,0,A\beta}/L$, which is interpreted here as a volume-averaged source rather than as transport through a single physical face of the control volume. The Aβ species are assumed to be spatially homogeneous within this effective extracellular/interstitial compartment.

Because tau fibrils are formed predominantly inside neurons whereas Aβ is predominantly extracellular, direct multiplication of intracellular tau and extracellular Aβ concentrations would not represent a physically meaningful reaction rate. The coupling is therefore expressed through an effective concentration $T_{tau}(t)$ of tau fibrillar material that is accessible to Aβ. $T_{tau}(t)$ is defined by diluting the accessible fraction of the intracellular tau fibril mass into the Aβ control volume:

$$T_{tau}(t) = f_{acc}\,\frac{V_{soma}}{L^3}\,M_{tau}(t) \quad (6)$$

where $0 \le f_{acc} \le 1$ is an effective accessibility factor. It represents, in a lumped manner, all processes that permit Aβ monomers to encounter tau fibrillar surfaces, including pathological overlap at synaptic or intracellular/extracellular interfaces; no specific transport pathway is assumed. The factor $V_{soma}/L^3$ converts the tau fibril mass concentration defined in the soma to an equivalent concentration in the Aβ control volume. Because tau acts catalytically, the interaction does not consume tau fibril mass.

Three pathways are considered for the formation of on-pathway Aβ nuclei and oligomers. Intrinsic primary nucleation is represented by

$$R_{1,A\beta} = k_{o1,A\beta}\,m_{A\beta}^2 \quad (7)$$

where the quadratic dependence corresponds to an effective primary-nucleation reaction order $n_c=2$, consistent with the Aβ42 kinetic analyses of (Mosconi et al., 2026) and (Dear et al., 2020). Although the reaction order may coincide with the critical nucleus size in simplified elementary nucleation models, it generally reflects the combined kinetics of multiple molecular steps and therefore does not imply that the critical Aβ nucleus contains exactly two monomers.

Homotypic Aβ secondary nucleation is represented as a fibril-surface-catalyzed process with $n_c = 2$,

$$R_{2,A\beta} = k_{o2,A\beta}\, m_{A\beta}^2 M_{A\beta} \tag{8}$$

so that the rate is proportional to Aβ fibril mass, used here as a surrogate for available fibril surface. In Eq. (8), $n_2 = 2$ is reported by (Mosconi et al., 2026). The baseline rate constant is instead the value specified in Table 2 and discussed below. Finally, the tau-mediated heterotypic primary-nucleation pathway is written in a form that can be calibrated directly from the seed-concentration-dependent primary-nucleation constants reported by (Mosconi et al., 2026):

$$R_{tau,A\beta} = \alpha_p\, T_{tau}\, m_{A\beta}^2 \tag{9}$$

where *p* denotes the tau fibril polymorph (PHF for the principal AD calculations and CTE for comparison), and $\alpha_p$ is a fold-dependent heterotypic nucleation coefficient. (Mosconi et al., 2026), Supplementary Table 1B, reports the fitted Aβ42 primary-nucleation constant $k_n$ at tau seed concentrations from 0 to 1 μM. Linear regression of $k_n$ against seed concentration gives slopes of 6.686 $M^{-1}$ $h^{-1}$ per μM for PHF and 16.010 $M^{-1}$ $h^{-1}$ per μM for CTE. After converting $M^{-1}$ to $10^{-6}$ $\mu M^{-1}$ and $h^{-1}$ to $s^{-1}$, these slopes give $\alpha_{PHF}$ = 1.86×10$^{-9}$ $\mu M^{-2}$ $s^{-1}$ and $\alpha_{CTE}$ = 4.45×10$^{-9}$ $\mu M^{-2}$ $s^{-1}$. Thus, the magnitude of the tau-catalyzed nucleation term is determined directly from the experimentally measured dependence of the fitted Aβ primary-nucleation constant on tau seed concentration, without introducing an additional independently assumed catalytic rate constant. (Mosconi et al., 2026) also observed positive cooperativity when varying Aβ42 concentration, but numerical Hill parameters $K_{0.5}$ and $n_H$ are not tabulated; they are therefore omitted from the baseline model rather than introduced as unconstrained parameters. Tau is not consumed by this catalytic term, and the Aβ elongation constant remains independent of tau.

Applying conservation of free Aβ monomers gives

$$\frac{dm_{A\beta}}{dt} = -R_{1,A\beta} - R_{2,A\beta} - \alpha_p T_\tau m_{A\beta}^2 + k_{d1,A\beta} S_{A\beta} + k_{d2,A\beta} S_{A\beta} M_{A\beta}{}^{eff}$$

$$-2k_{+,A\beta} m_{A\beta} P_{A\beta} - k_{\mathrm{conv},A\beta} m_{A\beta} S_{A\beta} - m_{A\beta} \frac{\ln 2}{T_{1/2,m,A\beta}} + \frac{q_{A0,A\beta}}{L} \tag{10}$$

The first three terms on the right-hand side of Eq. (10) represent Aβ monomer consumption through intrinsic primary nucleation, Aβ-fibril-surface secondary nucleation, and tau-catalyzed heterotypic nucleation, respectively. The third term is written explicitly as $\alpha_p T_{tau} m_{A\beta}^2$, where $\alpha_p$ is fold-specific because its value depends on the structural conformation of the tau fibril. Thus, the different catalytic activities of PHF and CTE tau fibrils measured in (Mosconi et al., 2026) are represented by different values of $\alpha_p$. The next two terms represent the release of Aβ monomers when oligomers dissociate, either spontaneously or through a process promoted by the Aβ fibril surface. The following two terms describe monomer consumption by fibril elongation and by conversion of on-pathway oligomers into elongation-competent fibrils. The penultimate term represents Aβ monomer clearance, and the final term is the effective volume-averaged Aβ production source. Tau-mediated nucleation is an additional pathway and does not replace spontaneous Aβ primary nucleation or homotypic Aβ secondary nucleation.

The secondary-nucleation term of Eq. (8) is written in unsaturated form, and the constant $k_{o2,A\beta}$ is taken from (Dear et al., 2020); the Materials and Methods of that study quote $k_2 \approx 10^4$ $M^{-2}s^{-1}$, that is $1.00\times10^{-8}$ $\mu M^{-2}s^{-1}$, for Aβ42 on the basis of the global fits of (Meisl et al., 2014) and (Cohen et al., 2013). Two independent checks support both the value and the functional form. First, (Dear et al., 2024) establish for Aβ42 and Aβ40 that $n_c$ and $n_2$ are both close to 2 over a wide range of monomer concentrations, so that the catalytic effect of the fibril surface is largely independent of monomer concentration; secondary nucleation is accordingly unsaturated throughout the regime in which these constants were determined, and (Mosconi et al., 2026) reach the same conclusion for Aβ42 self-aggregation, placing the saturation threshold above 6 μM. The present model operates at $m_{A\beta}$ of order $10^{-3}$ μM, three to four orders of magnitude below that threshold, so the unsaturated form of Eq. (8) is the appropriate limit rather than an approximation. Second, Eq. (1) of (Dear et al., 2020), evaluated with $n_2 = n_c = 2$, gives $k_2/k_n = 45$ $\mu M^{-1}$ for Aβ42, against 46.1 $\mu M^{-1}$ for the values adopted in Table 2, an agreement to within 2 % between constants taken from two different sources. An independent unsaturated determination is available from (Mosconi et al., 2026), Supplementary Table S1A, whose secondary-nucleation fit without the multistep term gives $k_2 = 1\times10^7$ $M^{-2}h^{-1}$, that is $2.78\times10^{-9}$ $\mu M^{-2}s^{-1}$. The two unsaturated estimates differ by a factor of 3.6 and bracket the range given in Table 2; the consequences of that spread for the accessibility threshold are reported in Section 3.

Homotypic Aβ secondary nucleation is catalyzed by the accessible fibril surface, not by the fibril mass as such. *In vitro* this distinction is less important when aggregate load is low and dispersed, so accessible surface approximately scales with fibril mass; in the interstitium, the model instead assumes that fibrillar Aβ consolidates into a compact plaque whose external surface grows as the two-thirds power of its volume. Retaining the unmodified form would therefore allow the surface term to grow without bound as the deposit consolidates, and by 33 years it exceeds intrinsic primary nucleation by four orders of magnitude, which suppresses every primary-nucleation channel including the tau-catalyzed one. The catalytic mass is accordingly restricted to a shell of thickness $\delta_{A\beta}$ at the plaque surface:

$$M_{A\beta}{}^{eff}(t) = M_{A\beta}(t)\min[1, \frac{3\delta_{A\beta}}{R_p(t)}], \quad R_p(t) = (\frac{3V_{ABP}(t)}{4\pi})^{1/3} \tag{11}$$

where VABP(t) is the plaque volume given by Eq. (51) and $R_p(t)$ is the radius of the equivalent sphere. The factor $\frac{3\delta_{A\beta}}{R_p}$ follows from the thin-shell approximation: the accessible shell volume, $4\pi R_p{}^2\delta_{A\beta}$, divided by the total plaque volume, $\frac{4\pi R_p{}^3}{3}$, equals $\frac{3\delta_{A\beta}}{R_p}$. For a small deposit, $R_p \leq 3\delta_{A\beta}$, the factor equals unity and the unmodified expression is recovered, so the calibration of $k_{o2,A\beta}$ against *in vitro* data is preserved exactly. At 33 years the plaque radius is 3.57 μm and the accessible fraction is $8.4\times10^{-3}$, a 119-fold reduction in effective catalyst. The quantity $M_{A\beta,eff}$ replaces $M_{A\beta}$ in the homotypic secondary-nucleation terms of Eqs. (10) and (12) only; it does not appear in the elongation, dissociation or mass-balance terms, so total Aβ fibril mass is unaffected.

Consistent with the Finke-Watzky model, an oligomer is interpreted as an activated monomer and is assumed to carry the same mass as a free monomer (Morris et al., 2008), (Iashchishyn et al., 2017). This represents a considerable simplification of the oligomer formation mechanism. Applying the conservation principle to the number of free on-pathway Aβ oligomers within the CV and normalizing the resulting equation by the CV's volume yields:

$$\frac{dS_{A\beta}}{dt} = R_{1,A\beta} + R_{2,A\beta} + \alpha_p T_\tau m_{A\beta}^2 - k_{d1,A\beta}S_{A\beta} - k_{d2,A\beta}S_{A\beta}M_{A\beta}{}^{eff}$$

$$-k_{conv,A\beta}m_{A\beta}S_{A\beta} - S_{A\beta}\frac{\ln 2}{T_{1/2,S,A\beta}} \tag{12}$$

The first three terms on the right-hand side of Eq. (12) are the gains of on-pathway Aβ oligomers/nuclei through intrinsic primary nucleation, homotypic fibril-surface secondary nucleation, and tau-mediated heterotypic primary nucleation. The third contribution is $\alpha_p T_\tau m_{A\beta}^2$ and is equal in magnitude and opposite in sign to the corresponding monomer-consumption term in Eq. (10). The next two terms describe oligomer loss through spontaneous and fibril-surface-catalyzed dissociation, the sixth term represents conversion of oligomers into elongation-competent fibrils, and the final term accounts for proteolytic degradation of free Aβ oligomers. In this coarse-grained description, tau-generated elongation-prone nuclei are assigned to the same on-pathway pool $S_{A\beta}$ as nuclei generated by the two homotypic mechanisms.

Applying the conservation principle to the number of Aβ fibrillar species of all lengths and normalizing by the CV's volume yields the governing equation for the total fibril concentration, regardless of length:

$$\frac{dP_{A\beta}}{dt} = k_{conv,A\beta} m_{A\beta}(t) S_{A\beta}(t) + k_{-,A\beta} M_{A\beta}(t) \tag{13}$$

The first term on the right-hand side of Eq. (13) is the counterpart of the oligomer-to-fibril conversion term in Eqs. (10) and (12) and represents formation of elongation-competent fibrils. This process is assumed to proceed through monomer addition to an on-pathway oligomer. Fibril degradation is neglected, whereas the second term represents generation of new fibrils through fragmentation of existing ones. Tau-mediated nucleation therefore affects $P_{A\beta}$ indirectly by increasing the population of on-pathway Aβ nuclei available for conversion; no direct tau-dependent term is added to fibril elongation or fragmentation.

The mass of fibrils within the CV is characterized by the total number of Aβ monomers incorporated into them. Applying conservation of this quantity and normalizing by the CV's volume yields the following equation:

$$\frac{dM_{A\beta}}{dt} = 2k_{+,A\beta} m_{A\beta}(t) P_{A\beta}(t) + 2k_{conv,A\beta} m_{A\beta}(t) S_{A\beta}(t) \tag{14}$$

In Eq. (14), the first term on the right-hand side describes fibril elongation through monomer addition, which can occur at either tip of the fibril. The second term describes the conversion of oligomers into fibrils, a process that involves monomer addition to the oligomer. The factor of 2 in this term reflects the incorporation of both a monomer and an oligomer into the fibril mass during this event.

Because the pathophysiological processes of AD unfold over decades (Mark and Brehmer, 2022), the long-term behavior of Aβ concentrations in the CV is effectively independent of initial conditions on this timescale, and all Aβ species can therefore be set to zero at the time of AD onset, $t = 0$. This leads to the following initial conditions for Eqs. (10) and (12)–(14):

At $t=0$: $m_{A\beta}=0$, $S_{A\beta}=0$, $P_{A\beta}=0$, $M_{A\beta}=0$ (15)

Table 1. Dependent variables used in the model.

| Symbol | Definition | Units |
|---|---|---|
| $m_{A\beta}(t)$ | Molar concentration of free Aβ monomers | μM |
| $M_{A\beta}(t)$ | Molar concentration of Aβ monomers incorporated into all fibrillar species, regardless of length (a measure of the total fibril mass expressed in monomer units) | μM |
| $P_{A\beta}(t)$ | Molar concentration of all Aβ fibrillar species, regardless of length (a measure of the total number of fibrils) | μM |
| $S_{A\beta}(t)$ | Molar concentration of free on-pathway Aβ oligomers | μM |
| $m_{tau}(t)$ | Molar concentration of free tau monomers | μM |
| $M_{tau}(t)$ | Molar concentration of tau monomers incorporated into all fibrillar species, regardless of length (a measure of the total fibril mass expressed in monomer units) | μM |
| $P_{tau}(t)$ | Molar concentration of all tau fibrillar species, regardless of length (a measure of the total number of fibrils) | μM |
| $S_{tau}(t)$ | Molar concentration of free on-pathway tau oligomers | μM |
| $T_{tau}$ | Effective concentration of tau fibrillar material accessible to Aβ in the extracellular/interstitial Aβ control volume (derived coupling variable) | μM |

Table 2. Model parameters.

| Symbol | Definition | Units | Value or range | Reference or estimation method | Value(s) used in computations |
| --- | --- | --- | --- | --- | --- |
| $a_{21}$ | Conversion factor from $\frac{\text{mol}}{\mu\text{m}^3}$ to μM | $\frac{\mu\text{M}\ \mu\text{m}^3}{\text{mol}}$ | | | $10^{21}$ |
| $c$ | Scaling constant in Eq. (59) relating combined accumulated neurotoxicity to biological age | $\mu\text{M}^{-1}$ | 8685 | | 8685 [a] |
| $k_{d1,A\beta}$ | Rate constant for the spontaneous dissociation of Aβ oligomers into monomers | $\text{s}^{-1}$ | $6.47\times10^{-7}$ [b] | (Dear et al., 2020) | 0, $6.47\times10^{-7}$ |
| $k_{d1,tau}$ | Rate constant for the spontaneous dissociation of tau oligomers into monomers | $\text{s}^{-1}$ | $10^{-3}$ – $10^{-4}$ [c] | (Shammas et al., 2015) | 0, $1.69\times10^{-4}$ |
| $k_{d2,A\beta}$ | Rate constant for the fibril-catalyzed dissociation of Aβ oligomers into monomers | $\mu\text{M}^{-1}\ \text{s}^{-1}$ | $1.99\times10^{-5}$ [d] | (Dear et al., 2020), see notes b and d | 0, $1.99\times10^{-5}$ |
| $k_{d2,tau}$ | Rate constant for the fibril-catalyzed dissociation of tau | $\mu\text{M}^{-1}\ \text{s}^{-1}$ | $5.6\times10^{-5}$ [c] | (Dear et al., 2024), (Rodriguez Camargo et al., 2021), | 0, $5.6\times10^{-5}$ |
| | oligomers into monomers | | | (Shammas et al., 2015) | |
| $k_{o1,A\beta}$ | Rate constant for intrinsic Aβ primary nucleation, $R_{1,A\beta} = k_{o1,A\beta} m_{A\beta}^2$ | $\mu M^{-1}\ s^{-1}$ | $\approx(1.7–3.1)\times10^{-10}$ [e] | (Mosconi et al., 2026), (Dear et al., 2020) | $2.17\times10^{-10}$ |
| $k_{o1,tau}$ | Effective first-order rate constant for intrinsic tau primary nucleation into oligomers | $s^{-1}$ | $2.78\times10^{-6}$ [f] | (Combs and Gamblin, 2012) | $2.78\times10^{-6}$ |
| $k_{o2,A\beta}$ | Rate constant for homotypic Aβ fibril-surface secondary nucleation, $R_{2,A\beta} = k_{o2,A\beta} m_{A\beta}^2 M_{A\beta}$ | $\mu M^{-2}\ s^{-1}$ | $2.78\times10^{-9}$–$1.00\times10^{-8}$ [e] | (Mosconi et al., 2026), (Dear et al., 2020), see note "e" | $1.00\times10^{-8}$ |
| $k_{o2,tau}$ | Effective rate constant for homotypic tau fibril-surface secondary nucleation | $\mu M^{-2}\ s^{-1}$ | $1.67\times10^{-4}$ [g] | (Rodriguez Camargo et al., 2021) | $1.67\times10^{-4}$ |
| $k_{+,A\beta}$ | Rate constant for Aβ fibril elongation by monomer addition | $\mu M^{-1}\ s^{-1}$ | $9.72\times10^{-2}$ (Mosconi et al., 2026); 3.06 (Dear et al., 2020) [h] | (Mosconi et al., 2026), (Dear et al., 2020) | $9.72\times10^{-2}$ |
| $k_{+,tau}$ | Rate constant for tau fibril elongation by monomer addition | $\mu M^{-1}\ s^{-1}$ | $4.2\times10^{-3}$ [i] | (Kundel et al., 2018) | $4.2\times10^{-3}$ |
| $k_{-,A\beta}$ | Rate constant for new Aβ fibril generation via fragmentation | $s^{-1}$ | $3.5 \times 10^{-11}$ | (Cook and Leyh, 2023) | 0 [j] |
| $k_{-,tau}$ | Rate constant for new tau fibril generation via fragmentation | $s^{-1}$ | $5.6 \times 10^{-11}$ [i] | (Kundel et al., 2018) | $5.6 \times 10^{-11}$ |
| $k_{conv,A\beta}$ | Rate constant for the conversion of Aβ oligomers into elongation-competent fibrils | $\mu M^{-1}\ s^{-1}$ | $1.83 \times 10^{-6}$ [k] | (Dear et al., 2020) | $1.83 \times 10^{-6}$ |
| $k_{conv,tau}$ | Rate constant for the conversion of tau oligomers into elongation-competent fibrils | $\mu M^{-1}\ s^{-1}$ | $8.9 \times 10^{-6}$ [l] | (Shammas et al., 2015) | $8.9 \times 10^{-6}$ |
| $L$ | Side length of the effective local extracellular/interstitial Aβ control volume; chosen as approximately half the mean inter-plaque spacing | μm | 50 | Estimated from the data presented in Fig. 1a of (Querol-Vilaseca et al., 2019) | 50 |
| $MW_{A\beta}$ | Molecular weight of an Aβ monomer, computed as the sum of atomic masses of all constituent atoms | $g\ mol^{-1}$ | $4.51 \times 10^{3}$ [m] | (Raskatov, 2019) | $4.51 \times 10^{3}$ |
| $MW_{tau}$ | Molecular weight of a tau monomer, computed as the sum of atomic masses of all constituent atoms | g mol$^{-1}$ | $3.7\times10^{4}$ - $4.6\times10^{4}$ [n] | (Mandelkow and Mandelkow, 2012) | $4.0\times10^{4}$ |
| $q_{A,0,A\beta}$ | Equivalent Aβ monomer production flux used to represent membrane production in the effective extracellular/interstitial control volume; $q_{A,0,A\beta}/L$ is the corresponding volume-averaged source term | μM μm s$^{-1}$ | $1.1\times10^{-5}$ [o] | (Raskatov, 2019) | $1.1\times10^{-5}$ |
| $q_{tau}$ | Rate at which tau monomers are supplied to the soma | mol s$^{-1}$ | $1.10\times10^{-23}$ | Eq. (50) | $1.10\times10^{-23}$ [p] |
| $t_{AD,onset}$ | Median age at AD onset | s | $1.58\times10^{9}$ [q] | (Grotz et al., 2015), (Hampel et al., 2021) | $1.58\times10^{9}$ |
| $t_{AD,f}$ | Mean survival time from AD onset, including the preclinical stage | s | $1.04\times10^{9}$ [r] | (Zanetti et al., 2009) | $1.04\times10^{9}$ |
| $T_{1/2,m,A\beta}$ | Half-life of Aβ monomers | s | $4.61\times10^{3}$ [s] | (Waters, 2010) | $4.61\times10^{3}$, $10^{20}$ (to represent $T_{1/2,m,A\beta}\to\infty$) |
| $T_{1/2,m,tau}$ | Half-life of tau monomers in the soma | s | $2.16\times10^{5}$ | (Poppek et al., 2006) | $2.16\times10^{5}$, $10^{20}$ (to represent $T_{1/2,m,tau}\to\infty$) |
| $T_{1/2,S,A\beta}$ | Half-life of free Aβ oligomers | s | $9.40\times10^{3}$ [t] | | $9.40\times10^{3}$, $10^{20}$ (to represent $T_{1/2,S,A\beta}\to\infty$) |
| $T_{1/2,S,tau}$ | Half-life of free tau oligomers in the soma | s | $8.38\times10^{5}$ [u] | | $8.38\times10^{5}$, $10^{20}$ (to represent $T_{1/2,S,tau}\to\infty$) |
| $V_{soma}$ | Volume of the soma | $\mu m^3$ | $4.19\times10^{3}$ [v] | | $4.19\times10^{3}$ |
| $\rho_{A\beta}$ | Density of the Aβ plaque | $g/\mu m^3$ | $1.35\times10^{-12}$ [v] | (Cavicchi et al., 2018) | $1.35\times10^{-12}$ |
| $\rho_{NFT}$ | Density of neurofibrillary tangles | $g/\mu m^3$ | $1.35\times10^{-12}$ [w] | (Cavicchi et al., 2018) | $1.35\times10^{-12}$ |
| $f_{acc}$ | Fraction of intracellular tau fibril mass accessible to Aβ | | 0–1 | Sensitivity parameter; no specific transport pathway is assumed | 0, 0.1, 1 |
| $\alpha_p$ | Fold-dependent kinetic coefficient describing the catalytic enhancement of Aβ primary nucleation by accessible tau fibrils | $\mu M^{-2}\ s^{-1}$ | PHF: $1.86\times10^{-9}$; CTE: $4.45\times10^{-9}$ [x] | Supplementary Table 1B in (Mosconi et al., 2026), linear regression | $1.86\times10^{-9}$ |
| $n_{c,A\beta}$ | Effective reaction order for intrinsic Aβ primary nucleation | | 2 | (Mosconi et al., 2026), (Dear et al., 2020) | 2 |
| $n_{2,A\beta}$ | Effective reaction order for homotypic Aβ fibril-surface secondary nucleation | | 2 | (Mosconi et al., 2026), (Dear et al., 2020) | 2 |
| $n_{c,tau}$ | Effective reaction order for intrinsic tau primary nucleation | | 1 | (Combs and Gamblin, 2012), (Rodriguez Camargo et al., 2021) | 1 |
| $n_{2,tau}$ | Effective reaction order for homotypic tau fibril-surface secondary nucleation | | 2 | (Rodriguez Camargo et al., 2021) | 2 |
| $\omega$ | Weighting factor in Eq. (57) | | $1.31\times10^{-5}$ | Eq. (58) | $1.31\times10^{-5}$ |
| $\Xi_{comb,crit}$ | Lethal threshold of combined accumulated neurotoxicity from Aβ and tau oligomers | μM·s | $6.17\times10^{4}$ | Section 2.6 | $6.17\times10^{4}$ |
| $n_{min,tau}$ | Minimum number of tau monomers in an elongation-competent fibril fragment; used in the fragmentation term of Eq. (3) | | $2\times10^{3}$ | Estimated from neurofibrillary-tangle filament morphology; see Section 2.1 | $2\times10^{3}$ |
| $\delta_{A\beta}$ | Thickness of the catalytically accessible surface shell of an Aβ plaque, used in the plaque-surface limitation of | μm | 0.010 | One Aβ fibril diameter; see Section 2.2 | 0.010 |

|  | homotypic secondary nucleation, Eq. (11) |  |  |  |  |
|---|---|---|---|---|---|

[a] The constant $c$ is calibrated such that the maximum accumulated neurotoxicity obtained numerically from the red curve in Fig. 5a, corresponding to $f_{acc} = 0$ ($\Xi_{comb,crit} = 6.17\times10^4$ μM·s), is assigned a biological age of 100 years.

[b] (Dear et al., 2020) reports a combined Aβ42 oligomer dissociation rate of $k_d$ = 0.36 h$^{-1}$ = 1.00×10$^{-4}$ s$^{-1}$, where $k_d = k_{d1,A\beta} + k_{d2,A\beta} M_{A\beta}$. In the kinetic model of (Dear et al., 2020), oligomers are treated explicitly as a separate species. The data do not uniquely identify $k_{d1,A\beta}$ and $k_{d2,A\beta}$ separately. The present pair $k_{d1,A\beta}$ = 6.47×10$^{-7}$ s$^{-1}$ and $k_{d2,A\beta}$ = 1.99×10$^{-5}$ μM$^{-1}$ s$^{-1}$ is not, however, an arbitrary decomposition of that total, because a second and independent constraint fixes the ratio of the two constants. Eq. (2) of (Dear et al., 2024), evaluated with $n_2 = n_c = 2$ for Aβ42, gives $k_{d2,A\beta}/k_{d1,A\beta} = (2/3)(k_{o2,A\beta}/k_{o1,A\beta})$ = 30.7 μM$^{-1}$, with no dependence on monomer concentration. Imposing this ratio together with the fitted total at the reference fibril-mass concentration $M_{A\beta}$ = 5 μM, $k_d$ = 1.00×10$^{-4}$ s$^{-1}$ determines both constants uniquely, giving $k_{d1,A\beta}$ = 6.47×10$^{-7}$ s$^{-1}$ and $k_{d2,A\beta}$ = 1.99×10$^{-5}$ μM$^{-1}$ s$^{-1}$. The total dissociation rate itself remains an experimental input.

[c] A consistent finding in oligomer-explicit aggregation studies is that a substantial fraction of oligomers dissociates back to monomer before conversion into growth-competent fibrils; at representative intracellular tau concentrations, the relevant kinetic comparison is $k_{conv,tau} m_{tau} < k_{d1,tau}$ (Michaels et al., 2020), (Shammas et al., 2015).

Single-molecule and NMR relaxation studies of amyloid oligomers report free-oligomer lifetimes ranging from milliseconds (fast, low-order transient self-association) to fast, dynamic dissociation on a sub-hour timescale. Two independent, directly relevant measurements bound free-oligomer dissociation between millisecond-scale (Aβ, tightly structured "dark-state" oligomers, $k_{d1}$ = 73 s$^{-1}$ (Fawzi et al., 2010)) and several-hour-or-shorter (α-synuclein, $k_{d1}$ > 5.6×10$^{-5}$ s$^{-1}$, i.e. > 0.2 h$^{-1}$ (Zurlo et al., 2021)) timescales. For tau, an order-of-magnitude estimate lying between these two bounds is adopted:

$$k_{d1,tau} \approx 10^{-3} - 10^{-4}\ \text{s}^{-1} \quad (16)$$

For wild-type K18 tau, (Shammas et al., 2015) directly fitted an oligomer dissociation rate of 0.61 $h^{-1}$ = $1.69\times10^{-4}$ $s^{-1}$. This tau-specific value lies within the adopted order-of-magnitude range and supports use of $10^{-4}$ $s^{-1}$ as the rounded representative value in the computations.

Direct measurements of fibril-surface-mediated dissociation of tau oligomers are not presently available. The corresponding rate constant $k_{d2,tau}$ was therefore estimated from the experimentally measured catalytic effect of tau fibril surfaces on oligomer formation. (Dear et al., 2024) showed that a fibril surface that catalyzes oligomer formation should also accelerate the reverse oligomer-dissociation process by approximately the same kinetic factor. In the notation used in that kinetic framework,

$$k_{d2,tau} \simeq \frac{2}{3}\frac{k_{2,tau}}{k_{n,tau}} m_{tau}^{n_{2,tau}-n_{c,tau}} k_{d1,tau} \tag{17}$$

Here, $k_{n,tau}$ denotes the intrinsic primary-nucleation rate constant for tau aggregation, $k_{2,tau}$ denotes the fibril-surface-mediated secondary-nucleation rate constant, $n_{c,tau}$ is the reaction order with respect to tau monomer concentration for primary nucleation, and $n_{2,tau}$ is the corresponding reaction order for secondary nucleation. These quantities are introduced here only because they are the parameters used in (Rodriguez Camargo et al., 2021) to characterize the relative catalytic strength of primary and fibril-surface-mediated tau nucleation. They are not additional parameters of the present model and do not enter its governing equations independently; they are used only to obtain a literature-based estimate of $k_{d2,tau}$.

(Rodriguez Camargo et al., 2021) reported for the $tau_{304-380,\mathrm{C322S}}$ fragment

$$k_{+,tau}k_{2,tau} = 6 \times 10^{11}\ \mathrm{M}^{-3}\mathrm{s}^{-2} \tag{18}$$

and

$$k_{+,tau}k_{n,\mathrm{tau}} < 10^{-2}\ \mathrm{M}^{-1}\mathrm{s}^{-2} \tag{19}$$

with $n_{2,tau} = 2$ and $n_{c,tau} = 1$. Because the same elongation constant $k_{+,tau}$ appears in both fitted products, it cancels when their ratio is taken, giving

$$\frac{k_{2,tau}}{k_{n,tau}} > 6 \times 10^{13}\ \mathrm{M}^{-2} \tag{20}$$

Substitution of $n_{2,tau} = 2$ and $n_{c,tau} = 1$ into the expression above gives

$$k_{d2,tau} \gtrsim \frac{2}{3}(6 \times 10^{13}\ \mathrm{M}^{-2}) m_{tau} k_{d1,tau} \tag{21}$$

For wild-type K18 tau, (Shammas et al., 2015) reported a free-oligomer dissociation rate of approximately

$$k_{d1,tau} = 0.61\ \mathrm{h}^{-1} = 1.69 \times 10^{-4}\ \mathrm{s}^{-1} \quad (22)$$

The secondary-nucleation mechanism fitted by (Rodriguez Camargo et al., 2021) is multistep and saturates. Writing the saturation constant of that mechanism as $K_{2,tau}$, which has units of $\mu M^2$ because the reaction order $n_{2,tau} = 2$, the unsaturated ratio in Eq. (21) must be divided by the factor $1 + m_{tau}^2/K_{2,tau}$; the concentration at which secondary nucleation is half saturated is $\sqrt{K_{2,tau}} \approx 30$ nM. The relevant monomer concentration is the free cytosolic tau available for aggregation, for which (Kundel et al., 2018) adopt 100 nM in human neurons, the great majority of intracellular tau being bound to microtubules. Evaluating Eq. (21) at $m_{tau} = 100$ nM $= 1\times10^{-7}$ M, where the saturation factor is 12.1, gives

$$\frac{2}{3}(6 \times 10^{13})(1 \times 10^{-7})/12.1 = 3.3 \times 10^{5}\ \mathrm{M}^{-1} = 0.33\ \mu\mathrm{M}^{-1} \quad (23)$$

Therefore,

$$k_{d2,tau} \approx 0.33\ k_{d1,tau} = 5.6 \times 10^{-5}\ \mu\mathrm{M}^{-1}\mathrm{s}^{-1} \quad (24)$$

Accordingly,

$$k_{d2,tau} = 5.6 \times 10^{-5}\ \mu\mathrm{M}^{-1}\mathrm{s}^{-1} \quad (25)$$

is adopted as the reference value. Because the estimate combines kinetic measurements obtained for different tau repeat-domain constructs and because the primary-nucleation contribution in (Rodriguez Camargo et al., 2021) is reported only as an upper bound, the resulting value should be regarded as an approximate, lower-bound-type estimate. The influence of $k_{d2,tau}$ on the model predictions is therefore examined by sensitivity analysis.

Eq. (20) also determines $k_{o2,tau}$, since $k_{2,tau} = (k_{2,tau}/k_{n,tau})\cdot k_{n,tau}$. With $k_{n,tau} = k_{o1,tau}$ this gives $k_{o2,tau} > 1.67\times10^{-4}\ \mu M^{-2}s^{-1}$, the lower-bound value adopted in Table 2. The same ratio constrains the secondary-nucleation rate constant and, through Eq. (17), the fibril-catalyzed dissociation constant $k_{d2,tau}$, so that $n_{c,tau}$, $n_{2,tau}$, $k_{o2,tau}$ and $k_{d2,tau}$ combine information from multiple experimental sources and should be regarded as approximate cross-construct estimates.

[d] The individual value of $k_{d2,A\beta}$ is not separately resolved by (Dear et al., 2020). The value $1.99\times10^{-5}$ $\mu M^{-1}\ s^{-1}$ ($= 7.15\times10^{-2}\ \mu M^{-1}\ h^{-1}$) follows from combining the fitted total rate with the catalytic ratio $k_{d2,A\beta}/k_{d1,A\beta} = 30.7\ \mu M^{-1}$ of (Dear et al., 2024), Eq. (2), as set out in note b. Together with $k_{d1,A\beta} = 6.47\times10^{-7}\ s^{-1}$ it reproduces the fitted total dissociation rate $k_d = 0.36\ h^{-1}$ at $M_{A\beta} = 5\ \mu M$ exactly. Because

the partition is now fixed by an independent relation rather than chosen, a separate sensitivity analysis of it is no longer required; the remaining experimental uncertainty resides in the fitted total $k_d$ .

[e] For intrinsic Aβ primary nucleation, (Mosconi et al., 2026) reports unseeded fitted values for the rate constant that corresponds to $k_{o1,A\beta}$ in the present notation. The values are $0.78\ \mathrm{M}^{-1}\mathrm{h}^{-1}$ for the PHF seed series and $0.60\ \mathrm{M}^{-1}\mathrm{h}^{-1}$ for the CTE seed series, equivalent to $2.17 \times 10^{-10}$ and $1.67 \times 10^{-10}\ \mu\mathrm{M}^{-1}\mathrm{s}^{-1}$, respectively. (Dear et al., 2020) gives the corresponding intrinsic primary-nucleation constant as $1.1 \times 10^{-6}\ \mu\mathrm{M}^{-1}\mathrm{h}^{-1} = 3.06 \times 10^{-10}\ \mu\mathrm{M}^{-1}\mathrm{s}^{-1}$. Accordingly, $k_{o1,A\beta} = 2.17 \times 10^{-10}\ \mu\mathrm{M}^{-1}\mathrm{s}^{-1}$, based on the PHF seed series of (Mosconi et al., 2026), is adopted as the baseline value, while the range 1.7–3.1 $\times 10^{-10}\ \mu\mathrm{M}^{-1}\mathrm{s}^{-1}$ is used for sensitivity analysis.

For homotypic Aβ fibril-surface secondary nucleation, the rate constant denoted $k_{o2,A\beta}$ in the present model is taken from (Dear et al., 2020); that study reports $3.6 \times 10^{-5}\ \mu\mathrm{M}^{-2}\mathrm{h}^{-1} = 1.00 \times 10^{-8}\ \mu\mathrm{M}^{-2}\mathrm{s}^{-1}$; this value is used as the baseline. (Mosconi et al., 2026), Supplementary Table 1A, gives a simplified secondary-nucleation comparison value equivalent to $2.78 \times 10^{-9}\ \mu\mathrm{M}^{-2}\mathrm{s}^{-1}$. This value is retained only for sensitivity analysis because the preferred tau-seeded fits in (Mosconi et al., 2026) employ a multistep saturated formulation rather than the unsaturated secondary-nucleation term used in the present model.

[f] A value of $0.01\ \mathrm{h}^{-1} = 2.78\times10^{-6}\ \mathrm{s}^{-1}$ was adopted from (Combs and Gamblin, 2012). An effective primary-nucleation order $n_{c,tau} = 1$ is used here, following the global kinetic fit of (Rodriguez Camargo et al., 2021); those authors note that the primary-nucleation rate is ill determined in their data and could be bounded only from above, so this order is the best-fitting rather than a tightly constrained value. The (Combs and Gamblin, 2012) constant is directly usable under this choice because the nucleation step A→B of the Finke–Watzky scheme they employ is first order in monomer, so that the tau primary-nucleation term in Eqs. (1) and (2) is written as $k_{o1,tau}\, m_{tau}$ with $k_{o1,tau}$ in $\mathrm{s}^{-1}$; note that the Finke–Watzky scheme is not adopted for the secondary step, whose order $n_{2,tau} = 2$ is taken from (Rodriguez Camargo et al., 2021). As an independent check, the upper limit $k_+k_n < 10^{-2}\ \mathrm{M}^{-1}\mathrm{s}^{-2}$ of (Rodriguez Camargo et al., 2021), combined with $k_{+,tau}$ from (Kundel et al., 2018), implies $k_{o1,tau} < 2.38\times10^{-6}\ \mathrm{s}^{-1}$, so the adopted value is approximately 17 % above that upper limit. This modest discrepancy is retained as a cross-construct inconsistency because the two determinations used different constructs and conditions. (Kundel et al., 2018) obtained $n_c = 2$ for heparin-induced full-length wild-type tau but $n_c = 1$ for the P301S variant under otherwise identical conditions, indicating that the apparent primary-nucleation order is sensitive to

sequence as well as to construct and solution conditions; their study is used here principally to constrain elongation and fragmentation.

[g] Estimated from the ratio $k_{2,tau}/k_{n,tau} > 6\times10^{13}$ M$^{-2}$ = 60 μM$^{-2}$ inferred from the fitted rate-constant products reported by (Rodriguez Camargo et al., 2021) for tau secondary nucleation with $n_{2,tau} = 2$ and $n_{c,tau} = 1$ (Eq. 19), combined with $k_{n,tau} = k_{o1,tau} = 2.78\times10^{-6}$ s$^{-1}$ (footnote f), giving $k_{o2,tau} > 1.67\times10^{-4}$ μM$^{-2}$s$^{-1}$. Because Eq. (20) provides only a lower bound, and because (Rodriguez Camargo et al., 2021) used a multistep secondary-nucleation mechanism that saturates with $\sqrt{K_M} \approx 30$ nM, the value obtained here should be interpreted as a low-monomer-limit, cross-construct estimate rather than as a uniquely determined *in vivo* constant. The present unsaturated term is therefore a coarse-grained approximation, and sensitivity to $k_{o2,tau}$ should be examined; explicit saturation is an important future refinement. The magnitude of this approximation was checked directly. In the regime the model occupies, $m_{tau} \approx 9\times10^{-5}$ μM, the saturation factor $1 + m_{tau}^2/K_{2,tau}$ is $1 + 2\times10^{-8}$, so the unsaturated term is the correct limit and saturated and unsaturated integrations agree to five significant figures. The approximation is therefore immaterial in Eqs. (1) and (2). The same factor is material in Eq. (23), where it is included: at the free cytosolic tau concentration of 100 nM adopted there it equals 12.1.

[h] For the Aβ fibril elongation rate constant $k_{+,A\beta}$, (Mosconi et al., 2026), Supplementary Table 1, reports $k_{+,A\beta} = 3.5\times10^{8}$ M$^{-1}$h$^{-1}$, equivalent to $9.72\times10^{-2}$ μM$^{-1}$s$^{-1}$. This value is adopted as the baseline because it is consistent with the kinetic framework used in (Mosconi et al., 2026) to calibrate tau-mediated Aβ nucleation. (Dear et al., 2020), Table S3, reports a larger Aβ elongation rate constant, $k_{+,A\beta} = 1.1\times10^{4}$ μM$^{-1}$h$^{-1}$ = 3.06 μM$^{-1}$s$^{-1}$, which is retained for sensitivity analysis. This sensitivity value is approximately 31.5 times larger than the baseline value and is used to assess the robustness of the model predictions to uncertainty in $k_{+,A\beta}$.

[i] (Kundel et al., 2018) used single-aggregate (SAVE) TIRF imaging to track the length distribution of full-length tau fibrils over aggregation times of up to two months, at monomer concentrations of 2 and 10 μM *in vitro*. The resulting length-versus-time data were fit to a three-step kinetic model comprising primary nucleation, saturated elongation, and fragmentation (their Eq. (2), following the analytical framework of (Knowles et al., 2009)). The fitted rate constants for wild-type (WT) tau (Fig. 2b in (Kundel et al., 2018)) are:

$k_{+,tau} = 4.2\times10^{-3}$ μM$^{-1}$s$^{-1}$ (equivalently $4.2\times10^{3}$ M$^{-1}$s$^{-1}$ as reported) (26)

$k_{-,tau} = 5.6\times10^{-11}$ s$^{-1}$ (27)

The (Kundel et al., 2018) measurements directly constrain fibril elongation and fragmentation and therefore support the fragmentation term in Eq. (3). They do not establish fragmentation as the predominant source of soluble oligomers. The isotope-labeling experiments of (Rodriguez Camargo et al., 2021) instead show that the oligomeric intermediates are generated predominantly from soluble monomers by fibril-surface secondary nucleation, which is why that process appears in Eqs. (1) and (2). The elongation constant $k_{+,tau}$ was fitted by (Kundel et al., 2018) within a saturated-elongation form, $2k_{+,tau} m_{tau} P_{tau}/(1 + m_{tau} /K_{E,tau})$, with $K_{E,tau}$ = 3.5 μM, the critical concentration above which elongation becomes independent of monomer concentration. This constant is unrelated to $K_{2,tau}$ of footnote c, which governs the saturation of secondary nucleation and has different units. The present model operates at $m_{tau} \approx 9\times10^{-5}$ μM, so the saturation factor is $1 + 3\times10^{-5}$ and the unsaturated expression used in Eqs. (1) and (4) is the appropriate limit; saturated and unsaturated integrations were verified to agree to five significant figures.

[j] The Aβ fibril fragmentation rate constant, $k_{-,A\beta}$, was estimated from the kinetic measurements of (Cook and Leyh, 2023), who fitted Aβ42 fibril-formation data using a nucleation–elongation–fragmentation model. For unliganded Aβ42, they reported a fragmentation rate constant of $(3.5 \pm 0.2) \times 10^{-11}\ \mathrm{s}^{-1}$. This literature value indicates that Aβ42 fibril fragmentation is a very slow process under quiescent conditions. In particular, the corresponding characteristic fragmentation timescale, $1/k_{-,A\beta}$, is approximately $9 \times 10^{2}$ years, which is far longer than the 33-year simulation period considered here. Fragmentation is therefore expected to have a negligible effect on the present results, and $k_{-,A\beta} = 0$ is used in the computations.

[k] (Dear et al., 2020) reports for Aβ42 a pseudo-first-order oligomer-to-fibril conversion rate $k_c$ = $3.3\times10^{-2}$ h$^{-1}$ at an initial monomer concentration $m_0$ = 5 μM. With the present conversion term $k_{conv,A\beta} m_{A\beta} S_{A\beta}$ and $n_{conv} = 1$, $k_{conv,A\beta} = k_c / m_0 = 6.6\times10^{-3}$ μM$^{-1}$ h$^{-1}$ = $1.83\times10^{-6}$ μM$^{-1}$ s$^{-1}$.

[l] Once a non-growth-competent tau oligomer forms, it may either dissociate back into monomers or undergo a structural conversion into a growth-competent fibrillar species. (Shammas et al., 2015) modeled this monomer-dependent oligomer-to-fibril conversion as a bimolecular process. The corresponding kinetic term has the same form as $k_{conv,tau} m_{tau} S_{tau}$ in the present model. For wild-type K18 tau, the fitted conversion rate constant reported in (Shammas et al., 2015) corresponds, in the present notation, to

$$k_{conv,tau} = (3.2 \pm 1.5) \times 10^{-2}\ \mu\mathrm{M}^{-1}\mathrm{h}^{-1} = (8.9 \pm 4.2) \times 10^{-6}\ \mu\mathrm{M}^{-1}\mathrm{s}^{-1} \quad (28)$$

The central value,

$$k_{conv,tau} = 8.9 \times 10^{-6}\ \mu\mathrm{M}^{-1}\mathrm{s}^{-1} \quad (29)$$

is adopted as the baseline estimate in the present calculations.

[m] $MW_{A\beta}$ corresponds to the molecular weight of Aβ42, as reported in (Raskatov, 2019).

[n] The value of $MW_{tau}$ varies according to the specific isoform involved (Mandelkow and Mandelkow, 2012).

[o] According to (Raskatov, 2019), Aβ42 monomers are produced in the human brain at a total rate of approximately

$$\dot{N}_{A\beta} = 0.33 \times 10^{-6}\ \mu\mathrm{mol}\ \mathrm{s}^{-1} \quad (30)$$

This estimate corresponds to approximately $3 \times 10^{11}$ Aβ42 molecules produced per second in the human brain. Dividing this value by the estimated brain volume,

$$V_{brain} = 1.5 \times 10^{15}\ \mu\mathrm{m}^3 \quad (31)$$

gives the average volumetric production rate

$$\dot{Q}_{A\beta} = \frac{\dot{N}_{A\beta}}{V_{brain}} = 2.2 \times 10^{-22}\ \mu\mathrm{mol}\ \mu\mathrm{m}^{-3}\mathrm{s}^{-1} \quad (32)$$

Since $1\ \mathrm{L} = 10^{15}\ \mu\mathrm{m}^3$, this is equivalent to the concentration source

$$Q_{A\beta} = 2.2 \times 10^{-7}\ \mu\mathrm{M}\ \mathrm{s}^{-1} \quad (33)$$

For the local cubic control volume of side length $L = 50\ \mu\mathrm{m}$, $V_{A\beta} = L^3 = 1.25 \times 10^5\ \mu\mathrm{m}^3$, the corresponding total Aβ42 production rate within the control volume is

$$\dot{N}_{A\beta,\mathrm{CV}} = \dot{Q}_{A\beta} L^3 = 2.75 \times 10^{-17}\ \mu\mathrm{mol}\ \mathrm{s}^{-1} \quad (34)$$

These values are the same as those used in the earlier formulation (Kuznetsov, 2026a).

Aβ is generated and released at neuronal membranes distributed throughout the local tissue (Dawkins et al., 2023), rather than entering the control volume through a single physical face. Accordingly, $Q_{A\beta}$ is interpreted as a volume-averaged membrane source. More generally, if $J_{A\beta}$ denotes the Aβ production/release flux per unit neuronal membrane area and $A_m$ is the total neuronal membrane area contained within the control volume, then

$$Q_{A\beta} = \frac{A_m}{V_{A\beta}} J_{A\beta} \quad (35)$$

To retain the source-term notation used in the present model and in earlier work, the equivalent parameter $q_{A,0,A\beta}$, with units of $\mu$M $\mu$m s$^{-1}$, is defined such that

$$\frac{q_{A,0,A\beta}}{L} = Q_{A\beta} \tag{36}$$

Therefore,

$$q_{A,0,A\beta} = Q_{A\beta} L = (2.2 \times 10^{-7}\ \mu\text{M s}^{-1})(50\ \mu\text{m}) = 1.1 \times 10^{-5}\ \mu\text{M}\ \mu\text{m s}^{-1} \tag{37}$$

Thus, $q_{A,0,A\beta} = 1.1 \times 10^{-5}\ \mu\text{M}\ \mu\text{m s}^{-1}$ is adopted as the baseline value. The term $q_{A,0,A\beta}/L$ in the governing equation should not be interpreted as transport through one face of the control volume; it represents the spatially averaged production and release of Aβ42 from neuronal membranes distributed throughout the local extracellular/interstitial tissue.

One conceptual point is worth emphasizing: $Q_{A\beta} = 2.2 \times 10^{-7}\ \mu\text{M s}^{-1}$ is the physical volumetric source; $q_{A,0,A\beta}$ is merely its equivalent representation in the existing $q/L$ formulation. Consequently, if $L$ is changed, $q_{A,0,A\beta}$ should change proportionally so that $q_{A,0,A\beta}/L = Q_{A\beta}$ remains fixed. This avoids reintroducing an unintended dependence of the actual Aβ production rate on control-volume size.

[p] Tau is synthesized predominantly in the neuronal soma and is normally transported along the axon. In AD, axonal transport is disrupted, leading to redistribution and accumulation of tau within the somatodendritic compartment (Mandelkow and Mandelkow, 2012), (Zempel et al., 2017). Using $\frac{V_{NFT,f}}{V_{soma}} = 0.081$ from (Kril et al., 2002) together with $t_{AD,f} = 1.04 \times 10^9$ s in Eq. (50), the corresponding effective supply rate of tau monomers to the soma is estimated as $1.10 \times 10^{-23}$ mol s$^{-1}$.

[q] The average age at AD diagnosis is 75 years (Grotz et al., 2015). However, because amyloid plaque accumulation is thought to begin roughly 25 years before dementia becomes clinically apparent (Hampel et al., 2021), the onset of pathological changes, $t_{AD,onset}$, is estimated to occur around age 50.

[r] Given an average post-diagnosis survival of 8 years (Zanetti et al., 2009), the total estimated disease timeline, $t_{AD,f}$, from the onset of Aβ deposition to death, spans 33 years: 25 years of preclinical progression plus 8 years following clinical diagnosis.

[s] Half-life of Aβ monomers, $T_{1/2,m,A\beta}$, is estimated to be 1.28 hours (Waters, 2010).

[t] Direct measurements of the proteolytic half-life of free Aβ oligomers in brain tissue are not available. However, (Takeda et al., 2013) measured the *in vivo* disappearance of soluble Aβ oligomers from brain interstitial fluid in awake APP/PS1 mice after acute inhibition of new Aβ production. Over an 8 h period, the concentration of low-molecular-weight Aβ42 oligomers decreased by 88.0%, leaving approximately 12% of the initial concentration. Approximating this decrease as first-order clearance gives

$$\lambda_{S,A\beta} = -\frac{\ln(0.12)}{8\ \text{h}} = 0.265\ \text{h}^{-1} \tag{38}$$

and hence

$$T_{1/2,S,A\beta} = \frac{\ln 2}{\lambda_{S,A\beta}} = 2.62\ \text{h} \approx 9.4 \times 10^3\ \text{s} \tag{39}$$

Accordingly, $T_{1/2,S,A\beta} = 2.6$ h is adopted as the reference value for the effective clearance half-life of free Aβ oligomers.

[u] A direct experimental measurement of the intracellular degradation half-life of free tau oligomers is not currently available. The closest quantitative *in vivo* estimate is provided by measurements of soluble tau turnover in a mouse model of tauopathy. (Yamada et al., 2015) reported a half-life of approximately 9.7 days for soluble tau, whereas insoluble tau turned over considerably more slowly, with a half-life of approximately 34.2 days. Because the variable $S_{tau}$ in the present model represents a soluble, nonfibrillar tau oligomer pool, the soluble-tau turnover time provides a more appropriate experimental proxy than the insoluble-tau value. Accordingly,

$$T_{1/2,S,tau} = 9.7\ \text{days} = 8.38 \times 10^5\ \text{s} \tag{40}$$

is adopted as the reference value. This estimate is consistent in order of magnitude with measurements in human iPSC-derived neurons, for which a tau half-life of $6.74 \pm 0.45$ days has been reported (Sato et al., 2018). Because neither study isolates tau oligomers specifically, $T_{1/2,S,tau}$ should be regarded as an experimentally constrained proxy for the turnover of the soluble tau pool rather than as a directly measured oligomer-specific half-life.

[v] The diameter of the soma is assumed to be 20 μm.

[w] Protein density is commonly approximated at 1.35 g/cm³, independent of the specific protein in question (Cavicchi et al., 2018).

[x] In the present model, the tau-catalyzed nucleation rate is written as

$$R_{tau,A\beta} = \alpha_p\, T_{tau}\, m_{A\beta}^2 \tag{41}$$

where $T_{tau}$ , given by Eq. (6), is the effective concentration of tau fibrillar material accessible to Aβ and $m_{A\beta}$ is the Aβ monomer concentration. Thus, $\alpha_p$ determines the increase in the Aβ primary-nucleation rate produced by a unit concentration of accessible tau fibrils.

It is called fold-dependent because different structural forms of tau fibrils have different catalytic activities. PHF and CTE tau fibrils expose different surface structures, and (Mosconi et al., 2026) found that CTE fibrils catalyze Aβ42 nucleation more strongly than PHF fibrils. Therefore,

$$\alpha_{CTE} > \alpha_{PHF} \tag{42}$$

Physically, $\alpha_p$ is an effective heterotypic nucleation coefficient: it incorporates the ability of a particular tau fibril fold to bind Aβ monomers and promote their conversion into new Aβ nuclei. It is not a tau consumption rate, because tau acts catalytically and is not consumed by the reaction.

The fold-dependent heterotypic coefficient $\alpha_p$ was calibrated directly from (Mosconi et al., 2026), Supplementary Table 1B. In (Mosconi et al., 2026), $k_n$ denotes the fitted effective Aβ42 primary-nucleation rate constant obtained for a given concentration of preformed tau fibril seeds. Thus, $k_n$ is not an additional parameter of the present model; rather, its dependence on tau seed concentration is used here to quantify the tau-induced increase in Aβ primary nucleation. Linear regression of the fitted $k_n$ values against tau seed concentration over the range 0–1 $\mu$M gives slopes of $6.686\ \mathrm{M^{-1}h^{-1}}$ per $\mu$M of PHF seeds and $16.010\ \mathrm{M^{-1}h^{-1}}$ per $\mu$M of CTE seeds. Because these slopes represent the increase in the effective Aβ primary-nucleation coefficient per unit tau fibril concentration, they correspond directly to the heterotypic coefficients $\alpha_{PHF}$ and $\alpha_{CTE}$ in the present model. Converting $\mathrm{M^{-1}}$ to $10^{-6}\ \mu\mathrm{M^{-1}}$ and $\mathrm{h^{-1}}$ to $\mathrm{s^{-1}}$ gives

$$\alpha_{PHF} = 1.86 \times 10^{-9}\ \mu\mathrm{M^{-2}s^{-1}} \tag{43}$$

and

$$\alpha_{CTE} = 4.45 \times 10^{-9}\ \mu\mathrm{M^{-2}s^{-1}} \tag{44}$$

These coefficients correspond to the fully accessible tau seed concentrations used in the *in vitro* experiments of (Mosconi et al., 2026). In the tissue-scale model, only a fraction of intracellular tau fibril mass is assumed to be accessible to extracellular/interstitial Aβ; this compartmental accessibility is represented separately by $f_{acc}$.

### 2.3. Scenario with infinite tau monomer and oligomer half-lives, $T_{1/2,m,tau} \to \infty$ and $T_{1/2,S,tau} \to \infty$

When degradation of tau monomers and oligomers is neglected, corresponding to infinite values of $T_{1/2,m,tau}$ and $T_{1/2,S,tau}$, tau introduced continuously into the soma is not removed from the system. The numerical solution (Fig. S1) shows that, after an initial transient, the free tau monomer concentration approaches a low quasi-steady value, whereas the oligomer concentration continues to increase slowly over the 33-year interval. Nevertheless, both soluble pools contain little mass compared with the fibrillar pool.

Under these conditions, the continuously supplied tau is ultimately transferred almost entirely into the fibrillar pool. Because $M_{tau}$ represents the concentration of tau monomer units incorporated into fibrils, conservation of tau mass implies that, when $m_{tau}$ and $S_{tau}$ have reached steady state, $M_{tau}$ increases at a constant rate equal to the net tau supply rate. Therefore, the fibril mass grows linearly with time. From Eq. (1), it follows that

$$M_{tau}\left(t\right)=\frac{q_{tau}}{V_{soma}}a_{21}t \tag{45}$$

### 2.4. Method for estimating the volume occupied by NFTs

The volume occupied by NFTs (Fig. 1) is evaluated from the total number of tau monomers incorporated into fibrillar aggregates within the neuronal soma, denoted by $N_{tau\,mon}$, at a given time *t*. Following the approach adapted from (Watzky et al., 2008), this quantity can be expressed as:

$$N_{tau\,mon}\left(t\right)=\frac{M_{tau}\left(t\right)}{a_{21}}V_{soma}N_{A} \tag{46}$$

where $N_{A}$ is Avogadro's number.

The same quantity, $N_{tau\,mon}$, can also be related to the physical volume of the NFTs using the formulation also described in (Watzky et al., 2008):

$$N_{tau\,mon}\left(t\right)=\frac{V_{NFT}\left(t\right)\rho_{NFT}}{MW_{tau}}N_{A} \tag{47}$$

where $MW_{tau}$ is the molecular weight of a tau monomer, $V_{NFT}\left(t\right)$ is the volume occupied by NFTs within the soma at time *t*, and $\rho_{NFT}$ is the density of the NFT material.

Equating the right-hand sides of Eqs. (46) and (47) and rearranging for the NFT volume gives

$$\frac{V_{NFT}(t)}{V_{soma}}=\frac{M_{tau}(t)}{a_{21}}\frac{MW_{tau}}{\rho_{NFT}} \tag{48}$$

Substitution of the expression for $M_{tau}$ from Eq. (45) into Eq. (48) then yields

$$\frac{V_{NFT}(t)}{V_{soma}}=\frac{MW_{tau}}{\rho_{NFT}}\frac{q_{tau}}{V_{soma}}t \tag{49}$$

Finally, evaluating Eq. (49) at $t=t_{AD,f}$ and solving for $q_{tau}$ gives

$$q_{tau}=\frac{V_{NFT}(t_{AD,f})}{V_{soma}}\frac{\rho_{NFT}}{MW_{tau}}\frac{V_{soma}}{t_{AD,f}} \tag{50}$$

Eq. (50) is particularly useful for parameter estimation because it provides a direct means of determining $q_{tau}$ from the prescribed NFT volume.

### 2.5. Modeling the growth of Aβ plaque volume

Aβ plaque growth is calculated using the framework developed in (Kuznetsov, 2024). Plaque growth (Fig. 1) is quantified in terms of the total number of Aβ monomers, $N_{A\beta mon}$, incorporated into the plaque by time *t*. Following the approach of (Watzky et al., 2008), $N_{A\beta mon}$ can be determined from the average concentration of Aβ monomers incorporated into the plaque:

$$N_{A\beta mon}(t)=2N_A\frac{L^3}{a_{21}}M_{A\beta}(t) \tag{51}$$

The factor of 2 on the right-hand side of Eq. (51) accounts for the assumption that the plaque develops between two neighboring neurons, with the membrane of each neuron supplying Aβ monomers at a rate of $q_{A,0,A\beta}$, thereby contributing to plaque growth (Fig. 1).

Alternatively, also following (Watzky et al., 2008), the same number of incorporated Aβ monomers can be related to the physical volume of a single plaque, $V_{ABP}$, according to

$$N_{A\beta mon}(t)=N_A\frac{V_{ABP}(t)\rho_{A\beta}}{MW_{A\beta}} \tag{52}$$

where $MW_{A\beta}$ is the average molecular weight of an Aβ monomer. Equating the right-hand sides of Eqs. (51) and (52) and solving for the plaque volume yields

$$V_{ABP}\left(t\right)=2\frac{MW_{A\beta}}{\rho_{A\beta}}\frac{L^3}{a_{21}}M_{A\beta}\left(t\right) \tag{53}$$

where $a_{21}=10^{21}\,\frac{\mu\text{M}\,\mu\text{m}^3}{\text{mol}}$ converts the units of $\frac{\text{mol}}{\mu\text{m}^3}$ to $\mu\text{M}$.

The fraction of the local control volume occupied by Aβ plaque material is

$$\frac{V_{ABP}\left(t\right)}{2L^3}=\frac{MW_{A\beta}}{\rho_{A\beta}}\frac{M_{A\beta}\left(t\right)}{a_{21}} \tag{54}$$

which provides a measure of the progressive accumulation of fibrillar Aβ within the extracellular/interstitial region.

### 2.6. Accumulated neurotoxicity as a surrogate measure of biological age

The neurotoxicity criterion is based on the time-integrated concentrations of soluble Aβ and tau oligomers, rather than on the instantaneous burdens of amyloid plaques or neurofibrillary tangles. This choice is motivated by substantial experimental and clinical evidence indicating that soluble oligomeric species of both proteins are major contributors to neuronal and synaptic dysfunction in AD. The importance of soluble Aβ was demonstrated early by (McLean et al., 1999); those authors found that soluble Aβ levels are elevated in AD and correlate much more closely with disease severity than does plaque burden. Subsequent studies have shown that Aβ oligomers disrupt synaptic signaling and interact directly with neuronal receptors, leading to progressive synaptic dysfunction and neuronal injury (Zhao et al., 2012), (Sengupta et al., 2016), (Chen et al., 2017), (Benilova et al., 2012). Soluble tau oligomers are likewise neurotoxic: extracellular tau oligomers impair long-term potentiation and memory (Fa et al., 2016), and both Aβ and tau oligomers can produce synaptic dysfunction through overlapping APP-dependent mechanisms (Puzzo et al., 2017). Studies of human synapses further demonstrate that Aβ and tau oligomers interact directly and that both contribute to synaptic pathology (Marcatti et al., 2022), while depletion of tau markedly attenuates Aβ-driven toxicity in human neurons, indicating that the neurotoxic effects of the two protein species are mechanistically interconnected (Ng et al., 2024). These findings support treating both soluble Aβ and tau oligomers as major contributors to cumulative neuronal injury. Because such damage develops over prolonged periods, the relevant neurotoxic burden is represented here by the time-integrated exposure to both oligomer species rather than by their concentrations at a single time point. Consistent with this temporal interpretation, (Bollack et al., 2024) showed that the rate of change in Aβ burden improves prediction of cognitive decline beyond a single static measurement, underscoring the importance of the trajectory of pathological protein accumulation over time.

Accumulated neurotoxicity of Aβ oligomers is therefore defined following (Kuznetsov, 2025b) as:

$$\Xi_{A\beta}(t) = \int_0^t S_{A\beta}(\hat{t})\, d\hat{t} \tag{55}$$

Similarly, the accumulated neurotoxicity associated with tau oligomers is defined following (Kuznetsov, 2025a) as

$$\Xi_{tau}(t) = \int_0^t S_{tau}(\hat{t})\, d\hat{t} \tag{56}$$

The total accumulated neurotoxicity arising from exposure to both Aβ and tau oligomers is then expressed as:

$$\Xi_{comb}(t) = \Xi_{A\beta}(t) + \omega \Xi_{tau}(t) \tag{57}$$

The relative neurotoxic potencies of Aβ and tau oligomers have not been established quantitatively under directly comparable experimental conditions. Equal weighting of their accumulated exposures is therefore adopted here as a first-order reference assumption rather than as an experimentally demonstrated equivalence. This choice is motivated by substantial evidence that both Aβ and tau oligomers independently impair synaptic plasticity and memory, converge on common synaptotoxic pathways, and can act synergistically when present together (Fa et al., 2016), (Puzzo et al., 2017), (Busche and Hyman, 2020), (Marcatti et al., 2022). Furthermore, tau depletion markedly attenuates Aβ-oligomer-induced neurodegeneration in human neurons (Ng et al., 2024), indicating that both species make substantial and interacting contributions to neuronal injury. In the absence of quantitative data that uniquely determine their relative weighting, equal contributions are therefore used as the baseline approximation, and the sensitivity of the predictions to this assumption is examined separately.

Setting the weighting factor such that these two factors contribute equally to $\Xi_{comb}$ yields:

$$\omega = \Xi_{A\beta}(t_{AD,f}) / \Xi_{tau}(t_{AD,f}) \tag{58}$$

The lethal threshold of combined accumulated neurotoxicity due to Aβ and tau oligomers was assumed to be equal to the value of $\Xi_{comb,crit}$ calculated for $f_{acc}=0$ under the limiting conditions of zero dissociation constants ($k_{d1,tau}$, $k_{d2,tau}$, $k_{d1,A\beta}$, $k_{d2,A\beta}$) and infinite monomer and oligomer half-lives, corresponding to zero degradation rates ($T_{1/2,m,tau}$, $T_{1/2,S,tau}$, $T_{1/2,m,A\beta}$, $T_{1/2,S,A\beta}$). This procedure gives a lethal-threshold value of $\Xi_{comb,crit}=6.17\times10^{4}$ μM·s as indicated by the red curve in Fig. 5a.

Following (Kuznetsov, 2026b), the biological age of neurons, hereafter referred to simply as biological age, is defined as follows:

$$\text{biological age }(\text{s})=\text{calendar age}+c\Xi_{comb}(t)=t_{AD,onset}+t+c\Xi_{comb}(t) \quad (59)$$

**3. Results**

The numerical results were first examined for the limiting case in which oligomer dissociation is neglected and the half-lives of tau and Aβ monomers and oligomers are taken to be infinite, corresponding to zero proteolytic degradation rates. This limiting case makes it possible to isolate the effect of tau-mediated Aβ nucleation from the effects of oligomer reversibility and protein clearance.

For the tau subsystem, the results are independent of the accessibility factor $f_{acc}$, because this parameter determines the fraction of intracellular tau fibrillar material accessible to Aβ but does not affect the governing equations for tau aggregation itself. Following a short initial transient, the concentration of free tau monomers, $m_{tau}$, falls steeply and thereafter decays slowly, reaching about $9\times10^{-5}$ μM after 33 years (Fig. S1a); the free monomer pool is held at this low level because monomers are captured at fibril ends as fast as they are supplied. The concentration of free on-pathway tau oligomers, $S_{tau}$, behaves quite differently: it rises steeply at first and then more slowly, but continues to increase throughout the simulated interval, reaching approximately 3.2 μM after 33 years (Fig. S1b). The curves obtained for the different values of $f_{acc}$ are indistinguishable.

The tau supply is transferred predominantly into the fibrillar pool. The concentration of tau fibrillar species of all lengths, $P_{tau}$, increases continuously but with a steadily decreasing slope, reaching approximately 3.6 μM by the end of the 33-year simulation (Fig. S2a). The concentration of tau monomers incorporated into fibrillar species, $M_{tau}$, is linear in time to within the resolution of the figure and reaches about $2.7\times10^{3}$ μM after 33 years (Fig. S2b). This linearity is a direct consequence of mass conservation: in the absence of degradation the entire tau supply is retained within the compartment, and the monomer and oligomer pools are negligible beside the fibrillar one, so that $M_{tau}$ grows at the constant rate set by $q_{tau}$. The analytical prediction for $M_{tau}$ is indistinguishable from the numerical solution.

Because $S_{tau}$ continues to grow rather than settling at a constant value, the accumulated neurotoxicity due to tau oligomers increases faster than linearly, reaching approximately $2.4\times10^{9}$ μM·s at the end of the simulated interval (Fig. S3a). The fraction of the soma occupied by tau fibrillar material follows $M_{tau}$ and therefore increases linearly, reaching about 0.08 after 33 years (Fig. S3b), which is the value used in Section 2.4 to calibrate the tau source term. The numerical and analytical solutions for the fibrillar volume fraction coincide closely. None of these quantities depends on $f_{acc}$, because this parameter couples tau fibrils to Aβ aggregation without feeding back into the tau aggregation equations.

The Aβ subsystem exhibits a markedly different dependence on $f_{acc}$. In the absence of dissociation and degradation, the free Aβ monomer concentration, $m_{A\beta}$, decreases with time after an initial buildup (Fig. 2a). Increasing $f_{acc}$ accelerates this decrease because a larger amount of tau fibrillar material becomes available to catalyze heterotypic Aβ primary nucleation. The difference between the curves becomes progressively larger with time, and the smallest free-monomer concentration is obtained for the largest value of $f_{acc}$.

The opposite behavior is observed for the concentration of free on-pathway Aβ oligomers, $S_{A\beta}$ (Fig. 2b). The oligomer concentration increases throughout the simulated disease interval, and both its magnitude and its rate of increase rise substantially with increasing $f_{acc}$. At approximately 33 years, $S_{A\beta}$ is about $3.3\times10^{-5}$ μM for $f_{acc} = 0$, $5.1\times10^{-5}$ μM for $f_{acc} = 0.1$, and $1.5\times10^{-4}$ μM for $f_{acc} = 1$. Once homotypic secondary nucleation is restricted to the accessible plaque surface, the tau-catalyzed pathway is no longer negligible, but the amplification it produces is modest: a 54 % increase in the soluble oligomer concentration requires that a tenth of the somatic tau fibrillar material be accessible to interstitial Aβ, and complete accessibility increases it 4.6-fold.

The effect of tau-mediated nucleation is also apparent in the concentration of Aβ fibrillar species, $P_{A\beta}$ (Fig. 3a). $P_{A\beta}$ increases continuously for all values of $f_{acc}$, but the increase becomes progressively faster as $f_{acc}$ increases. By the end of the simulation, $P_{A\beta}$ reaches approximately $3.6\times10^{-4}$ μM for $f_{acc} = 0$, $4.2\times10^{-4}$ μM for $f_{acc} = 0.1$, and $7.0\times10^{-4}$ μM for $f_{acc} = 1$. Increasing the accessibility of tau fibrils therefore increases the number of Aβ fibrillar species, consistent with the generation of additional elongation-competent nuclei through the tau-catalyzed heterotypic pathway, but the increase is only a factor of two even when the whole of the tau fibrillar mass is made accessible.

In contrast, the total Aβ fibril mass, represented by $M_{A\beta}$, is nearly independent of $f_{acc}$ and increases approximately linearly with time (Fig. 3b). All three curves remain essentially superimposed and reach approximately 225–230 $\mu$M after 33 years. Thus, tau-mediated nucleation increases the number of Aβ fibrillar species but has little effect on their total mass under these limiting conditions. In the absence of degradation, the total amount of Aβ accumulated over a sufficiently long period is controlled primarily by the continuous monomer supply. Increasing $f_{acc}$ changes how this material is distributed among soluble and fibrillar species and increases the number of fibrils, but does not substantially alter the total fibrillar mass.

The pronounced effect of $f_{acc}$ on soluble Aβ oligomers is reflected directly in accumulated Aβ neurotoxicity (Fig. 4a). For $f_{acc} = 0$, accumulated Aβ oligomer exposure reaches approximately $3.1\times10^4$ μM·s after 33 years. Raising $f_{acc}$ to 0.1 increases this value to approximately $4.4\times10^4$ μM·s, and complete accessibility ($f_{acc}$ = 1) to approximately $1.2\times10^5$ μM·s. Inverting this relation gives the central quantitative result of the present model: a 50 % increase in accumulated Aβ oligomer exposure requires $f_{acc} \approx 0.12$, and a doubling requires $f_{acc} \approx 0.27$. Tau-templated heterotypic nucleation can therefore materially amplify the Aβ oligomer burden only if of order 10 % or more of the somatic tau fibrillar material becomes accessible to the interstitium. Values of $f_{acc}$ in this range are difficult to reconcile with tangles that remain enclosed within an intact neuronal membrane, and point instead to neuronal lysis and the formation of extracellular ghost tangles as the configuration in which the mechanism could operate at the scale considered here. It should be noted that $f_{acc}$ = 1 corresponds to $T_{tau}$ = 90.5 μM, some ninety times the highest tau seed concentration at which $\alpha_p$ was calibrated, so that case is reported as an extrapolated limiting case rather than as a prediction. The sensitivity of these thresholds to $k_{o2,A\beta}$ was examined over the range given in Table 2: with the lower value, $2.78\times10^{-9}$ μM$^{-2}$s$^{-1}$, the same amplifications require $f_{acc} \approx 0.035$ and 0.078, while pairing the upper value with the elongation constant of (Dear et al., 2020) rather than that of (Mosconi et al., 2026) gives 0.144 and 0.322. The threshold for a resolvable effect therefore lies between about 0.035 and 0.15 for every self-consistent choice of the Aβ parameter set, so the conclusion drawn here does not depend on which determination of $k_{o2,A\beta}$ is adopted.

The plaque volume normalized by the combined reference volume $2L^3$, by contrast, is essentially independent of $f_{acc}$ (Fig. 4b). The three curves are nearly indistinguishable and increase approximately linearly, reaching about $7.6\times10^{-4}$ by the end of the simulation. This is consistent with the fibril-mass results of Fig. 3b: tau-catalyzed heterotypic nucleation produces more Aβ fibrils but does not appreciably change the total mass incorporated into the fibrillar pool. Tau-mediated Aβ nucleation therefore increases accumulated Aβ neurotoxicity and the number of fibrillar species without producing a comparable increase in plaque volume, although the magnitude of that increase is modest at accessibilities below about 0.1.

The combined accumulated neurotoxicity due to Aβ and tau oligomers shows the same dependence on $f_{acc}$ (Fig. 5a). For $f_{acc}$ = 0 it reaches approximately $6.2\times10^4$ μM·s after 33 years, rising to approximately $7.4\times10^4$ μM·s for $f_{acc}$ = 0.1 and $1.5\times10^5$ μM·s for $f_{acc}$ = 1. Because the tau oligomer contribution is itself independent of $f_{acc}$, the whole of this increase comes from the additional Aβ oligomer burden generated through tau-catalyzed heterotypic nucleation. The weighting factor ω is calibrated so that the two compartments contribute equally at $f_{acc}$ = 0, so the value at that point is by construction twice the accumulated Aβ exposure reported above.

The corresponding biological-age predictions are shown in Fig. 5b. At early times biological age remains close to calendar age for all three values of $f_{acc}$, and the curves diverge progressively as accumulated oligomer exposure grows. For $f_{acc}$ = 0 the biological age of 100 years is reached at a calendar age of 83.0 years, which is the calibration point of Eq. (58) and is therefore fixed by construction rather than predicted. Tau-mediated Aβ nucleation advances that point: the same biological age is reached at 81.1 years for $f_{acc}$ = 0.1 and at 73.8 years for $f_{acc}$ = 1, and the biological ages at the end of the simulation are 100.1, 103.6 and 124.1 years respectively. The nine-year advance obtained at complete accessibility is substantial, but it requires the whole of the somatic tau fibrillar mass to be available to the interstitium, and the corresponding value of $T_{tau}$ lies far outside the range over which $\alpha_p$ was calibrated.

The importance of oligomer dissociation becomes apparent when physiologically relevant dissociation constants are used while proteolytic degradation remains absent (Fig. 6). Under these conditions the combined accumulated neurotoxicity is reduced by approximately three orders of magnitude relative to the corresponding case without dissociation: after 33 years it reaches only about 67, 83 and 168 μM·s for $f_{acc}$ = 0, 0.1 and 1 respectively (Fig. 6a), against $6.2\times10^4$ to $1.5\times10^5$ μM·s in Fig. 5a, a suppression by a factor of about 930 at $f_{acc}$ = 0. The relative dependence on $f_{acc}$ survives, the burden still rising by a factor of 2.5 between $f_{acc}$ = 0 and 1, but the absolute magnitude is far smaller because dissociation returns a large fraction of the oligomers to the monomer pool before they can contribute to accumulated exposure.

As a consequence, biological age is indistinguishable from calendar age once oligomer dissociation is included (Fig. 6b). The combined exposure remains between two and three orders of magnitude below the threshold $\Xi_{comb,crit}$, so that even for $f_{acc}$ = 1 the predicted biological age at the end of the simulation is 83.1 years against a calendar age of 83.0 years. The three curves are superimposed on the scale of the figure, in marked contrast to the separation obtained in Fig. 5. Reversible oligomer formation therefore suppresses the long-term accumulation of neurotoxic oligomer exposure almost completely at these parameter values.

A qualitatively similar suppression is obtained when oligomer dissociation is neglected but physiologically relevant finite half-lives of Aβ and tau monomers and oligomers are imposed (Fig. 7). Proteolytic clearance reduces the combined accumulated neurotoxicity to approximately $8.8\times10^{2}$ μM·s after 33 years, a factor of about seventy below the limiting case of Fig. 5a, and the residual dependence on $f_{acc}$ is weak: the values for $f_{acc}$ = 0, 0.1 and 1 are 879, 880 and 882 μM·s respectively (Fig. 7a). The reason is that clearance acts far more strongly on the tau oligomer pool, which dominates $\Xi_{comb}$, than the tau-catalyzed pathway acts on the Aβ pool. The accumulated Aβ exposure taken alone does retain a strong relative dependence on $f_{acc}$, rising from $6.6\times10^{-3}$ to 2.5 μM·s between $f_{acc}$ = 0 and 1, but it is negligible beside the tau contribution once clearance is active. Physiological protein turnover therefore opposes the aggregation processes that drive the model toward an AD-like state, and impairment of the degradation machinery appears to be a prerequisite for sustained pathological accumulation of neurotoxic oligomers. The result is insensitive to the tau monomer half-life: reducing it from 9.7 to 2.5 days changes $\Xi_{comb}$ by 0.03 % and leaves the predicted biological age unchanged, because the free tau monomer pool is too small for its clearance to remove appreciable mass and the suppression is governed almost entirely by the oligomer half-life $T_{1/2,S,tau}$.

The corresponding biological-age curves show that physiological protein turnover removes, rather than merely delays, the effect of tau-mediated Aβ nucleation on this measure (Fig. 7b). For all three values of $f_{acc}$ the predicted biological age at the end of the simulation is 83.3 years against a calendar age of 83.0 years, so that the three curves are superimposed on the scale of the figure and the 100-year threshold is not approached. Two features of the model account for this. First, the combined accumulated exposure remains about seventy times below $\Xi_{comb,crit}$, so that the additive term $c\Xi_{comb}$ contributes only about a quarter of a year to the biological age. Second, under these conditions $\Xi_{comb}$ is dominated by the tau oligomer contribution, which does not depend on $f_{acc}$; the tau-catalyzed increase in Aβ oligomer exposure, although large in relative terms, is negligible beside it. Once proteolytic clearance is active, the coupling is therefore too weak to advance the biological age at any accessibility.

Taken together, the results reveal a clear distinction between fibrillar burden and oligomer-mediated neurotoxicity. Increasing the fraction of tau fibrillar material accessible to Aβ has no effect on tau aggregation itself and, in the absence of degradation, only a minor effect on total Aβ fibril mass or plaque volume, yet it increases the number of Aβ fibrillar species, the concentration of soluble Aβ oligomers and the accumulated Aβ neurotoxicity by factors of 2.0, 4.6 and 3.8 respectively at complete accessibility.

The model therefore predicts that tau-catalyzed heterotypic Aβ nucleation can increase neurotoxic oligomer exposure without a proportional increase in visible plaque burden, but that it does so appreciably only when a large fraction of the somatic tau fibrillar material is accessible to the interstitium. Both oligomer dissociation and proteolytic degradation suppress the effect further, by three orders of magnitude and by a factor of about seventy respectively, indicating that intact protein turnover can limit the persistent accumulation of potentially toxic Aβ and tau species. Conversely, impairment of proteostatic mechanisms permits sustained oligomer accumulation and amplifies cumulative neurotoxicity by roughly two orders of magnitude, supporting the view that deterioration of protein homeostasis may be an important factor in the development and progression of AD.

**4. Discussion, limitations, and future directions**

The central result of the present model is that tau-mediated heterotypic nucleation can alter the distribution of Aβ among monomeric, oligomeric, and fibrillar states without producing a comparable increase in total Aβ fibril mass. Increasing the fraction of intracellular tau fibrillar material accessible to Aβ, $f_{acc}$, accelerates depletion of free Aβ monomers and increases the concentration of soluble Aβ oligomers. At the same time, it produces a larger number of Aβ fibrillar species, $P_{A\beta}$, while leaving the total concentration of Aβ incorporated into fibrils, $M_{A\beta}$, nearly unchanged in the limiting case without dissociation or degradation. This distinction between fibril number and fibril mass is important because it suggests that the principal effect of tau-mediated nucleation may be to change the manner in which Aβ mass is distributed among aggregates rather than simply to increase total plaque burden.

The plaque-surface limitation introduced in Eq. (11) is a geometric idealization. It treats the fibrillar Aβ within the control volume as a single compact sphere and the catalytic zone as a shell one fibril diameter thick, whereas real plaques are irregular, porous and multiple, and their accessible area is correspondingly larger than the equivalent-sphere estimate. The predicted accessibility threshold, $f_{acc} \approx 0.035$–$0.15$ for a resolvable effect, depending on the adopted value of $k_{o2,A\beta}$, should therefore be viewed as a lower-bound estimate under the compact-plaque assumption: a more porous deposit would retain more catalytic surface, suppress the tau-catalyzed channel further and raise the threshold, while a more compact one would lower it. The direction of the conclusion is nevertheless robust, because the unmodified form, in which the catalytic term scales with the whole fibril mass, makes homotypic secondary nucleation exceed intrinsic primary nucleation by four orders of magnitude within a day of simulated time and renders the tau-catalyzed pathway undetectable at any accessibility. A quantitative test of $\alpha_p$ in the regime where it was measured, that is, a closed system with a finite initial monomer load and no continuous source, in which the fibril mass never reaches the levels attained here, remains the most direct route to validating the heterotypic rate constant rather than merely calibrating it, and is reported in Section S2 of the Supplemental Materials.

This prediction is directly consistent with the experimental findings of (Mosconi et al., 2026). In their *in vitro* cross-seeding experiments, both PHF and CTE tau fibrils accelerated Aβ42 aggregation in a concentration-dependent manner, with CTE fibrils displaying the stronger catalytic effect. Kinetic analysis showed that tau fibrils promote heterotypic Aβ42 nucleation, and transmission electron microscopy demonstrated a reduction in Aβ42 fibril length under cross-seeding conditions. (Mosconi et al., 2026) interpreted the latter observation as a consequence of enhanced nucleation producing a larger number of growing fibrils and thereby reducing the amount of free Aβ42 monomer available for elongation. The present model reproduces this qualitative mechanism: increasing $f_{acc}$ lowers $m_{A\beta}$, increases $P_{A\beta}$, and leaves $M_{A\beta}$ comparatively insensitive to tau accessibility. Thus, the mathematical results provide a mechanistic interpretation of the experimentally observed shortening of Aβ fibrils in the presence of tau seeds.

The ratio $\frac{M_{A\beta}}{P_{A\beta}}$ provides a model-based measure of the mean number of Aβ monomer units incorporated per fibril and therefore a proxy for average fibril length. At the end of the simulated interval, $M_{A\beta}$ is approximately 225–230 $\mu$M for all investigated values of $f_{acc}$, whereas $P_{A\beta}$ increases from approximately $3.6 \times 10^{-4}$ $\mu$M for $f_{acc} = 0$ to approximately $4.2 \times 10^{-4}$ and $7.0 \times 10^{-4}$ $\mu$M for $f_{acc} =$ 0.1 and 1, respectively. Consequently, the calculated mean fibril mass per fibril decreases substantially as tau-mediated nucleation becomes stronger. Although the model does not resolve the complete fibril-length distribution, this prediction is qualitatively supported by the direct fibril-length measurements of (Mosconi et al., 2026); shorter Aβ42 fibrils were observed there after tau cross-seeding.

The predicted increase in Aβ oligomer concentration with increasing tau accessibility is also consistent with the broader experimental evidence that enhanced nucleation can increase the population of transient oligomeric intermediates. (Michaels et al., 2020), using combined kinetic measurements and oligomer quantification during Aβ42 aggregation, demonstrated that oligomers constitute a dynamic population formed along aggregation pathways and that only a minority ultimately converts into fibrils, whereas most dissociate back to monomers. The present model therefore predicts that tau-mediated acceleration of nucleation can affect not only fibril number and length but also the transient concentration and lifetime-integrated exposure to soluble Aβ oligomers.

This result is particularly relevant because experimental and human neuropathological studies indicate that soluble Aβ species are more closely related to neuronal dysfunction than bulk plaque burden. (McLean et al., 1999) measured soluble and insoluble Aβ in human AD brain and found an approximately threefold increase in soluble Aβ in AD, with soluble Aβ correlating strongly with markers of disease severity. In contrast, insoluble Aβ differentiated AD from control tissue but did not correlate with disease severity or plaque number. These observations are consistent with the present prediction that systems with similar total Aβ fibril masses or plaque burdens can nevertheless experience substantially different soluble oligomer exposures.

The model further predicts that increasing $f_{acc}$ can produce a several-fold increase in accumulated Aβ oligomer neurotoxicity even when the final plaque volume changes little. This separation between plaque burden and oligomer-mediated toxicity has experimental precedent. In addition to the human tissue results of (McLean et al., 1999), numerous experiments have shown that soluble Aβ oligomers can directly impair synaptic function. More importantly for the present coupled Aβ–tau framework, soluble tau oligomers are also independently neurotoxic. (Fa et al., 2016) demonstrated that extracellular tau oligomers rapidly impair hippocampal long-term potentiation and memory, whereas corresponding monomeric tau preparations did not produce these effects. Thus, the use of soluble Aβ and tau oligomer exposure, rather than plaque and tangle mass alone, as the basis of the model's neurotoxicity criterion is consistent with experimental observations.

There is also direct experimental evidence that Aβ and tau oligomers do not act independently. (Fa et al., 2016) found that concentrations of Aβ and tau oligomers that were individually below the threshold required to impair long-term potentiation or memory produced marked impairment when applied together. (Puzzo et al., 2017) similarly showed that extracellular Aβ and tau oligomers impair long-term potentiation and memory and that these effects depend on amyloid precursor protein. In experiments on isolated human synaptic terminals, (Marcatti et al., 2022) further demonstrated concentration-dependent interactions between Aβ and tau oligomers: high Aβ oligomer concentrations promoted recruitment of tau oligomers to synapses, while increasing tau oligomer concentrations could displace synaptic Aβ oligomers; Aβ oligomers also increased the tau-induced suppression of synaptic potentiation. These observations provide experimental support for representing the cumulative toxicity of Aβ and tau oligomers jointly, although they also indicate that their interaction may be nonlinear and disease-stage dependent.

The experimental study of (Ng et al., 2024) provides an additional connection between the two protein systems. Using human iPSC-derived cortical neurons, they found that Aβ exposure caused neuronal hyperactivity, impaired mitochondrial axonal transport, and neurodegeneration, and that these adverse effects were attenuated by tau depletion. Thus, experimental data support the view that Aβ- and tau-dependent toxicity are mechanistically interconnected rather than simply additive independent processes. This is consistent with the strong influence of tau–Aβ coupling predicted by the present model, although the current formulation represents this interaction at the aggregation-kinetic level rather than through downstream cellular signaling pathways.

The physical possibility that intracellular tau can become accessible to extracellular or synaptic Aβ is also supported experimentally, although the magnitude of this process remains poorly quantified. (Fein et al., 2008) demonstrated co-localization of Aβ and tau pathology in synaptosomes isolated from AD brain, providing evidence that the two proteins can occupy the same diseased synaptic compartments. (Guo et al., 2006) showed experimentally by Western blotting and ELISA that Aβ and tau can form stable soluble complexes, demonstrating direct molecular interaction between the two proteins. More recently, (Mosconi et al., 2026) demonstrated direct interaction between Aβ42 and disease-relevant tau fibrils by co-immunoprecipitation and showed that tau fibrils catalyze Aβ42 nucleation. These findings support the basic premise of the coupling term used here.

Furthermore, extracellular availability of tau does not require catastrophic neuronal disruption. (Pooler et al., 2013) showed experimentally that neuronal activity stimulates calcium-dependent release of endogenous tau from mature cortical neurons. (Yamada et al., 2014), using *in vivo* microdialysis, subsequently demonstrated that increasing neuronal activity rapidly raises extracellular tau concentrations in living mouse brain. These studies provide biological support for the assumption that at least some intracellular tau can become available in extracellular or synaptic environments where interaction with Aβ is possible. Nevertheless, they do not provide a quantitative estimate of the fraction of intracellular fibrillar tau that is accessible to Aβ. Consequently, $f_{acc}$ remains a phenomenological sensitivity parameter rather than an experimentally measured quantity.

Another result of the model is the strong protective effect of oligomer dissociation. Introducing physiologically relevant dissociation constants reduces the predicted accumulated neurotoxicity by approximately three orders of magnitude relative to the irreversible-oligomer case and keeps biological age close to calendar age. This prediction is consistent with direct experimental evidence that Aβ oligomers are highly dynamic rather than obligatorily committed to fibril formation. (Michaels et al., 2020) found that most experimentally detected Aβ42 oligomers dissociate back into monomers and that only a minority convert into fibrils. (Dear et al., 2024) further demonstrated experimentally that the dissociation of Aβ oligomers formed on fibril surfaces is itself strongly catalyzed by those surfaces. Thus, the strong sensitivity of the predicted oligomer burden to dissociation is consistent with measured Aβ aggregation kinetics.

*In vivo* measurements likewise support the importance of oligomer clearance. (Takeda et al., 2013) detected soluble low- and high-molecular-weight Aβ oligomers in the interstitial fluid of awake APP/PS1 mice and showed that these populations declined at different rates after acute inhibition of new Aβ production. Low-molecular-weight Aβ42 oligomers decreased by approximately 88% over 8 h, whereas high-molecular-weight oligomers decreased substantially more slowly. These experiments demonstrate that soluble Aβ oligomers undergo appreciable *in vivo* turnover and that their persistence depends strongly on oligomeric state. The present finding that oligomer lifetime is a major determinant of cumulative neurotoxicity is therefore qualitatively consistent with experimental observations, although the single effective oligomer pool used here cannot distinguish the heterogeneous lifetimes of different oligomer sizes and conformations.

The predicted importance of finite tau turnover is likewise biologically plausible. (Yamada et al., 2015) measured tau turnover *in vivo* in a mouse model of tauopathy and found that soluble and insoluble tau populations have distinct turnover times, with insoluble tau being substantially longer lived. Such observations support inclusion of finite tau clearance in the model, but they also emphasize that a single first-order half-life is necessarily a coarse-grained representation of multiple degradation and clearance pathways. Importantly, the present simulations show that physiologically effective turnover of Aβ and tau species strongly suppresses accumulated oligomer neurotoxicity, whereas impairment of degradation allows these species to persist and their cumulative toxic burden to increase markedly. This result raises the possibility that failure or substantial deterioration of the protein degradation machinery may be a prerequisite for the sustained accumulation of Aβ and tau species required for the development of pronounced AD pathology. In this interpretation, impaired proteostasis is not merely a consequence of protein aggregation but can act as a permissive condition that allows otherwise reversible and continuously cleared aggregation intermediates to accumulate over decades. Because the model can still generate appreciable toxicity under sufficiently strong tau–Aβ coupling when degradation remains active, however, this should be regarded as a model-derived mechanistic hypothesis rather than evidence that complete degradation failure is strictly required for AD.

The fold dependence incorporated through $\alpha_p$ is another feature with direct experimental support. (Mosconi et al., 2026) found that PHF and CTE tau fibrils differ substantially in their ability to catalyze Aβ42 nucleation, with CTE fibrils displaying the greater catalytic activity. They also observed structural effects on the Aβ42 fibrils generated in the presence of the different tau polymorphs and showed that PHF- and CTE-mediated cross-seeding increased Aβ42 toxicity both in SH-SY5Y cells and in transgenic *Caenorhabditis elegans*, again with fold-dependent differences. The use of different values of $\alpha_p$ for different tau folds is therefore based on experimentally observed kinetic and biological differences rather than being introduced solely as a modeling assumption.

Despite this experimental support, several important simplifications remain. First, the model treats accessible tau fibril mass as a surrogate for catalytic surface availability. In reality, catalytic activity is likely to depend on fibril length, surface area, structural defects, polymorph, and the number of exposed reactive sites. The experimentally observed fold dependence reported by (Mosconi et al., 2026) demonstrates that equal fibril masses need not possess equal catalytic activity. Future formulations should therefore distinguish total fibril mass from catalytic surface area and, where data become available, represent the distribution of reactive sites explicitly.

Second, the present moment-based model does not resolve the complete Aβ fibril-length distribution. It tracks $P_{A\beta}$, the concentration of fibrillar species, and $M_{A\beta}$, the fibrillar mass concentration. Their ratio can therefore be used only as a measure of mean fibril size. Section S2 provides a preliminary comparison of $M_{A\beta}/P_{A\beta}$ with microscopy: the model predicts a mean Aβ fibril length of 3.5 µm, compared with approximately 500 nm for self-aggregated and 250 nm for seeded Aβ42. (Mosconi et al., 2026) also observed the same direction of change, with tau cross-seeding reducing Aβ42 fibril length. Future work should compare predicted and measured fibril-length distributions across multiple tau-seeding conditions. Simultaneous experimental measurements of free Aβ monomers, oligomers, fibril number, fibril mass, and fibril-length distributions would provide a stringent test of the predicted redistribution mechanism.

Third, coupling is one-way in the present formulation: tau fibrils alter Aβ nucleation, whereas Aβ does not influence the tau aggregation equations. This assumption is useful for isolating the mechanism demonstrated by (Mosconi et al., 2026), but experiments indicate more extensive bidirectional interactions. (Guo et al., 2006) showed direct Aβ–tau complex formation, (Fa et al., 2016) and (Puzzo et al., 2017) demonstrated functional interactions between Aβ and tau oligomers, and (Ng et al., 2024) showed that several Aβ-induced neuronal phenotypes are tau-dependent. Future models should therefore investigate reciprocal Aβ-to-tau coupling, including possible positive-feedback mechanisms.

The accumulated-neurotoxicity criterion also requires qualification. Experimental studies strongly support neurotoxic roles for soluble oligomers of both proteins: (Fa et al., 2016) showed that tau oligomers impair synaptic plasticity and memory, while their combined application with Aβ oligomers at individually subthreshold concentrations produced pronounced functional deficits. (Marcatti et al., 2022) demonstrated direct AβO–TauO interactions at synapses, and (Puzzo et al., 2017) showed common APP-dependent mechanisms for oligomer-induced impairment. These observations justify including both Aβ and tau oligomer burdens in the neurotoxicity measure. They do not, however, establish that equal integrated concentrations of Aβ and tau oligomers produce equal neuronal damage. The equal weighting adopted here should therefore be regarded as a neutral first-order assumption necessitated by the absence of quantitative comparative dose–response data. Future experiments comparing Aβ and tau oligomers under identical conditions could be used to determine separate weighting coefficients and to test whether their combined toxicity is additive, synergistic, concentration dependent, or disease-stage dependent.

A still stronger limitation concerns the conversion of combined accumulated oligomer toxicity into biological age. The present model assumes a linear relationship between cumulative oligomer exposure and biological age after calibration of the proportionality coefficient. No experimental study has established such a linear relationship. Consequently, biological age in this model should be interpreted as a model-derived index of cumulative molecular injury, rather than as a validated physiological measure of aging. The absolute predicted biological-age values, particularly values exceeding a normal human lifespan, should therefore be viewed as measures of relative accumulated damage rather than literal ages. Future validation would require longitudinal measurements of Aβ and tau oligomer exposure together with neuronal, imaging, and cognitive outcomes. Such data could determine whether the dose–damage relationship is linear or instead contains thresholds, saturation, repair, synergy, or other nonlinear effects.

Finally, the model assumes spatially homogeneous concentrations within the intracellular tau and extracellular/interstitial Aβ control volumes. Experimental evidence indicates that Aβ–tau interactions can occur at highly localized sites. (Fein et al., 2008) demonstrated co-localization of Aβ and tau pathology within diseased synaptic terminals, while (Marcatti et al., 2022) showed direct concentration-dependent competition and recruitment of Aβ and tau oligomers at isolated human synapses. These findings suggest that local concentrations at synapses may differ substantially from volume-averaged concentrations. A spatially resolved reaction–diffusion model incorporating tau release, extracellular transport, Aβ diffusion, and localized synaptic interactions would therefore be an important extension of the present framework.

Taken together, the model results are consistent with several independent experimental observations: tau fibrils catalyze Aβ42 nucleation and produce shorter Aβ fibrils (Mosconi et al., 2026); soluble Aβ burden is more closely associated with disease severity than bulk plaque burden (McLean et al., 1999); soluble tau and Aβ oligomers independently and jointly impair synaptic plasticity and memory (Fa et al., 2016), (Puzzo et al., 2017); the two oligomer species interact directly at human synapses (Marcatti et al., 2022); and Aβ-induced neuronal dysfunction is strongly tau-dependent (Ng et al., 2024). The distinctive mechanistic prediction of the present framework is therefore not simply that tau increases Aβ plaque formation. Rather, tau-mediated heterotypic nucleation can redistribute Aβ into a larger population of shorter fibrils, deplete the free-monomer pool, and markedly alter the transient oligomer burden while producing a much smaller change in total fibrillar mass. The simulations further suggest that efficient protein turnover provides an important barrier to this pathological accumulation and that failure or substantial impairment of the protein degradation machinery may be a prerequisite for the sustained oligomer burden associated with AD. This prediction is already qualitatively supported by the tau-seeding experiments of (Mosconi et al., 2026) and provides a specific quantitative hypothesis for future experimental testing.

**Author Contribution Statement**

AVK is the sole author of this paper.

**Funding Data**

National Science Foundation (Grant No. DMS-2451660; Funder ID: 10.13039/100000146).

Alexander von Humboldt Foundation through the Humboldt Research Award (Funder ID: 10.13039/100005156).

**Conflict of Interest**

The author declares no competing interests.

**Data and Code Availability Statement**

No new experimental data were generated in this modeling study. All numerical inputs and their literature sources are reported in Table 2 and the Supplemental Materials. The experimental data used for calibration are available from Mosconi et al. (2026) and its associated repository (https://doi.org/10.5281/zenodo.18957210). The MATLAB scripts and input parameter files used to generate the reported results are available from the corresponding author on reasonable request.

**Ethics Statement**

None.

**Figure captions**

Fig. 1. Schematic of the coupled intracellular tau and extracellular/interstitial Aβ model. (a) Tau aggregation is modeled within a representative neuronal soma of volume $V_{soma}$, whereas Aβ aggregation is described within a surrounding extracellular/interstitial control volume $V_{A\beta} = L^3$. Tau monomers, oligomers, fibrillar species, and total fibril mass are represented by $m_{tau}$, $S_{tau}$, $P_{tau}$, and $M_{tau}$, respectively; the corresponding Aβ variables are $m_{A\beta}$, $S_{A\beta}$, $P_{A\beta}$, and $M_{A\beta}$. A fraction $f_{acc}$ of intracellular tau fibrillar material is assumed to become accessible to Aβ, giving the effective tau fibril concentration $T_{tau}$ in the Aβ compartment. (b) Model coupling between the two compartments. Accessible tau fibrils catalyze heterotypic primary nucleation of Aβ, thereby increasing the formation of Aβ oligomeric nuclei without consumption of tau fibril mass. Aβ also undergoes intrinsic primary nucleation, fibril-surface secondary nucleation, oligomer-to-fibril conversion, fibril elongation, fragmentation, dissociation, and proteolytic clearance.

Fig. 2. (a) Molar concentration of free Aβ monomers as a function of time, $m_{A\beta}(t)$. (b) Molar concentration of free on-pathway Aβ oligomers as a function of time, $S_{A\beta}(t)$. Results are shown for zero dissociation constants ($k_{d1,tau}$, $k_{d2,tau}$, $k_{d1,A\beta}$, $k_{d2,A\beta}$) and infinite half-lives ($T_{1/2,m,tau}$, $T_{1/2,S,tau}$, $T_{1/2,m,A\beta}$, $T_{1/2,S,A\beta}$).

Fig. 3. (a) Molar concentration of Aβ fibrillar species of varying length as a function of time, $P_{A\beta}(t)$. (b) Molar concentration of Aβ monomers incorporated into fibrillar species of varying length as a function of time, $M_{A\beta}(t)$. Results are shown for zero dissociation constants ($k_{d1,tau}$, $k_{d2,tau}$, $k_{d1,A\beta}$, $k_{d2,A\beta}$) and infinite half-lives ($T_{1/2,m,tau}$, $T_{1/2,S,tau}$, $T_{1/2,m,A\beta}$, $T_{1/2,S,A\beta}$).

Fig. 4. (a) Accumulated neurotoxicity caused by Aβ oligomers as a function of time, $\Xi_{A\beta}(t)$. (b) Normalized Aβ plaque volume, $V_{ABP}/(2L^3)$, as a function of time. Results are shown for zero dissociation constants ($k_{d1,tau}$, $k_{d2,tau}$, $k_{d1,A\beta}$, $k_{d2,A\beta}$) and infinite half-lives ($T_{1/2,m,tau}$, $T_{1/2,S,tau}$, $T_{1/2,m,A\beta}$, $T_{1/2,S,A\beta}$).

Fig. 5. (a) Combined accumulated neurotoxicity due to Aβ and tau oligomers as a function of time, $\Xi_{comb}(t)$. (b) Biological age as a function of calendar age. Results are shown for zero dissociation constants ($k_{d1,tau}$, $k_{d2,tau}$, $k_{d1,A\beta}$, $k_{d2,A\beta}$) and infinite half-lives ($T_{1/2,m,tau}$, $T_{1/2,S,tau}$, $T_{1/2,m,A\beta}$, $T_{1/2,S,A\beta}$).

Fig. 6. (a) Combined accumulated neurotoxicity due to Aβ and tau oligomers as a function of time, $\Xi_{comb}(t)$. (b) Biological age as a function of calendar age. Results are shown for physiologically relevant values of dissociation constants ( $k_{d1,tau}$ , $k_{d2,tau}$ , $k_{d1,A\beta}$ , $k_{d2,A\beta}$ ) but infinite half-lives ( $T_{1/2,m,tau}$ , $T_{1/2,S,tau}$ , $T_{1/2,m,A\beta}$ , $T_{1/2,S,A\beta}$ ).

Fig. 7. (a) Combined accumulated neurotoxicity due to Aβ and tau oligomers as a function of time, $\Xi_{comb}(t)$. (b) Biological age as a function of calendar age. Results are shown for zero dissociation constants ( $k_{d1,tau}$ , $k_{d2,tau}$ , $k_{d1,A\beta}$ , $k_{d2,A\beta}$ ) but physiologically relevant values of half-lives ( $T_{1/2,m,tau}$ , $T_{1/2,S,tau}$ , $T_{1/2,m,A\beta}$ , $T_{1/2,S,A\beta}$ ).

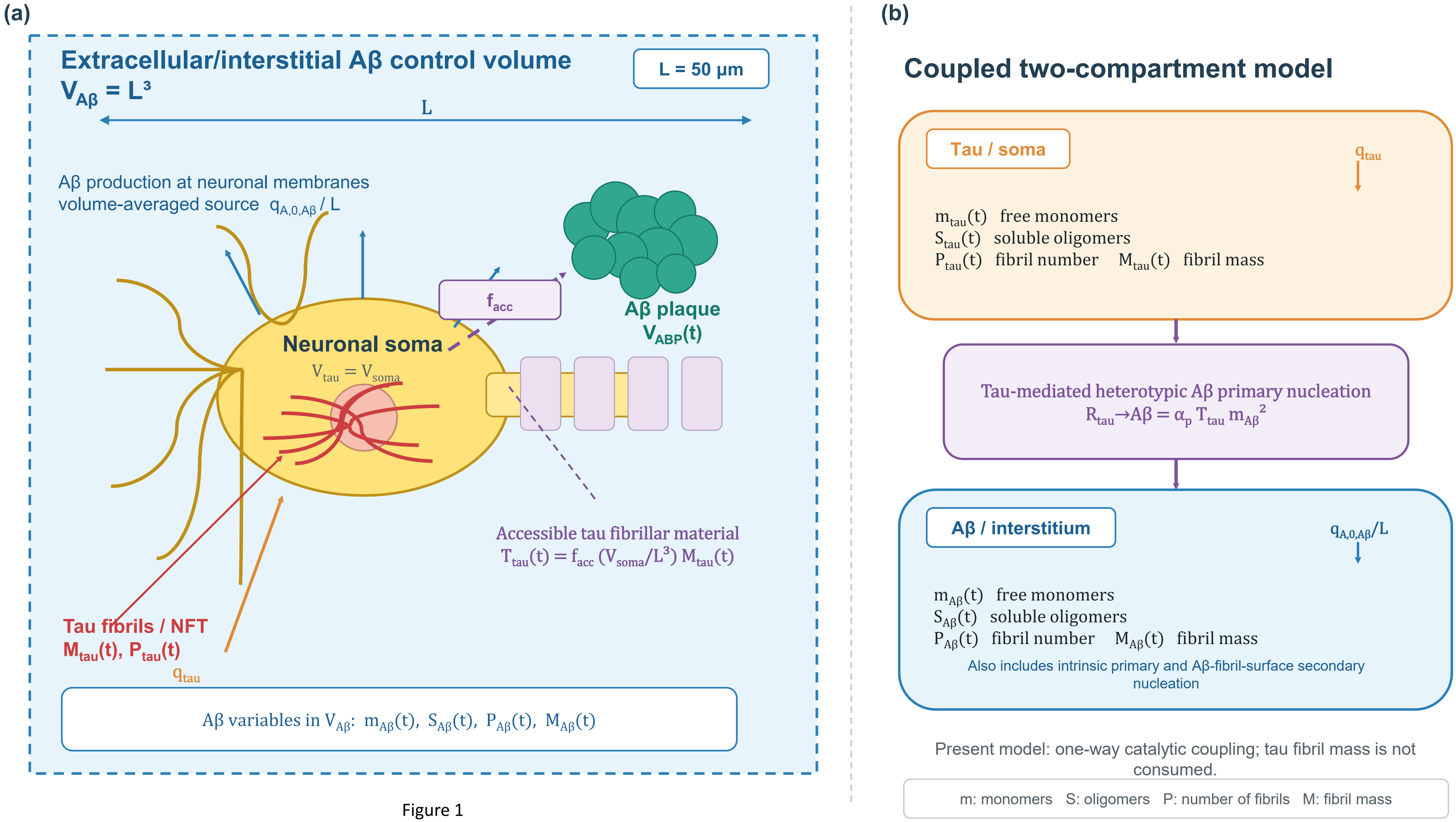

(a)

Extracellular/interstitial Aβ control volume

$V_{A\beta} = L^3$

L = 50 μm

L

Aβ production at neuronal membranes
volume-averaged source $q_{A,0,A\beta} / L$

$f_{acc}$

Aβ plaque
$V_{ABP}(t)$

Neuronal soma
$V_{tau} = V_{soma}$

Accessible tau fibrillar material
$T_{tau}(t) = f_{acc} (V_{soma}/L^3) M_{tau}(t)$

Tau fibrils / NFT
$M_{tau}(t)$, $P_{tau}(t)$

$q_{tau}$

Aβ variables in $V_{A\beta}$: $m_{A\beta}(t)$, $S_{A\beta}(t)$, $P_{A\beta}(t)$, $M_{A\beta}(t)$

(b)

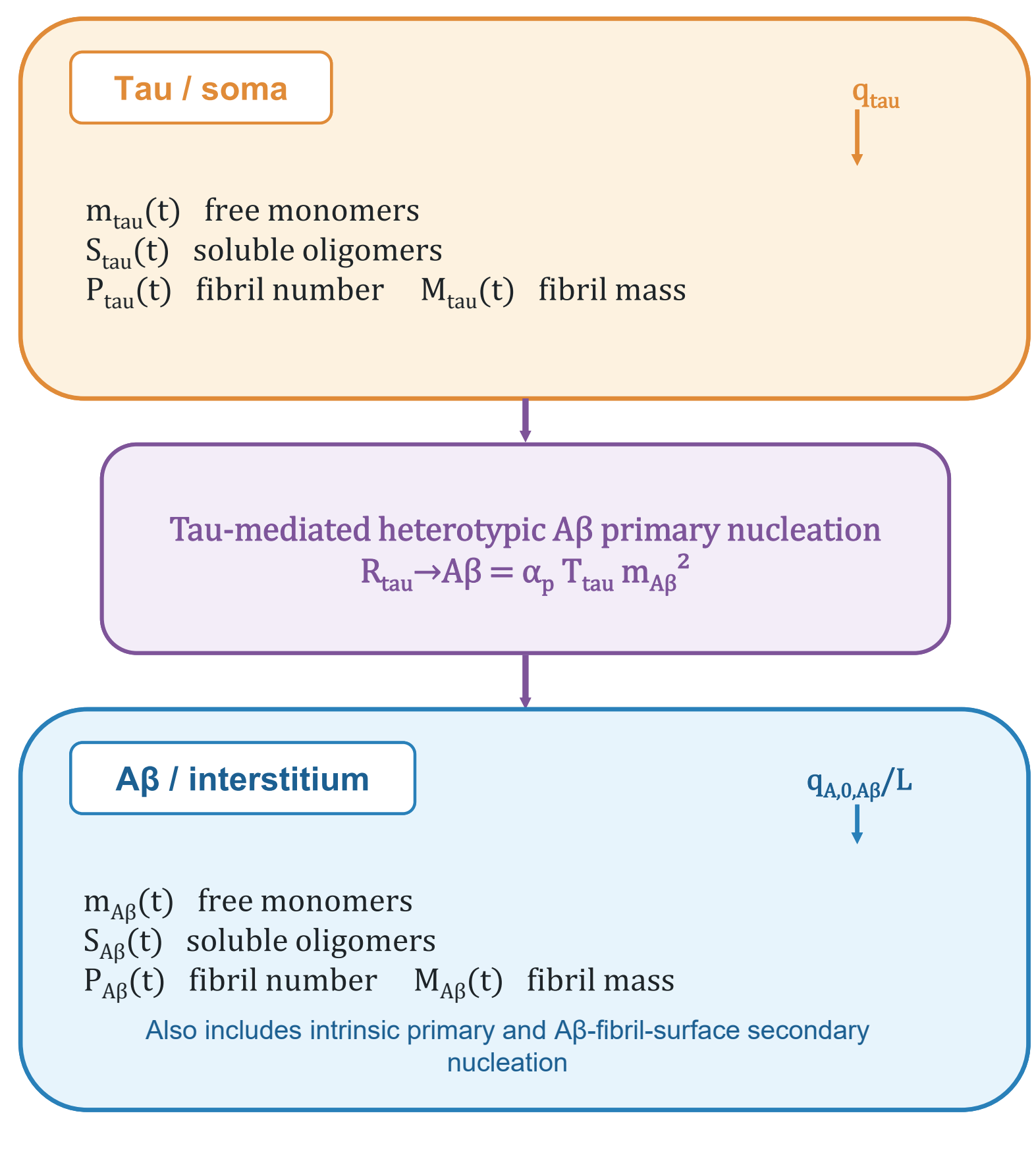

Figure 1

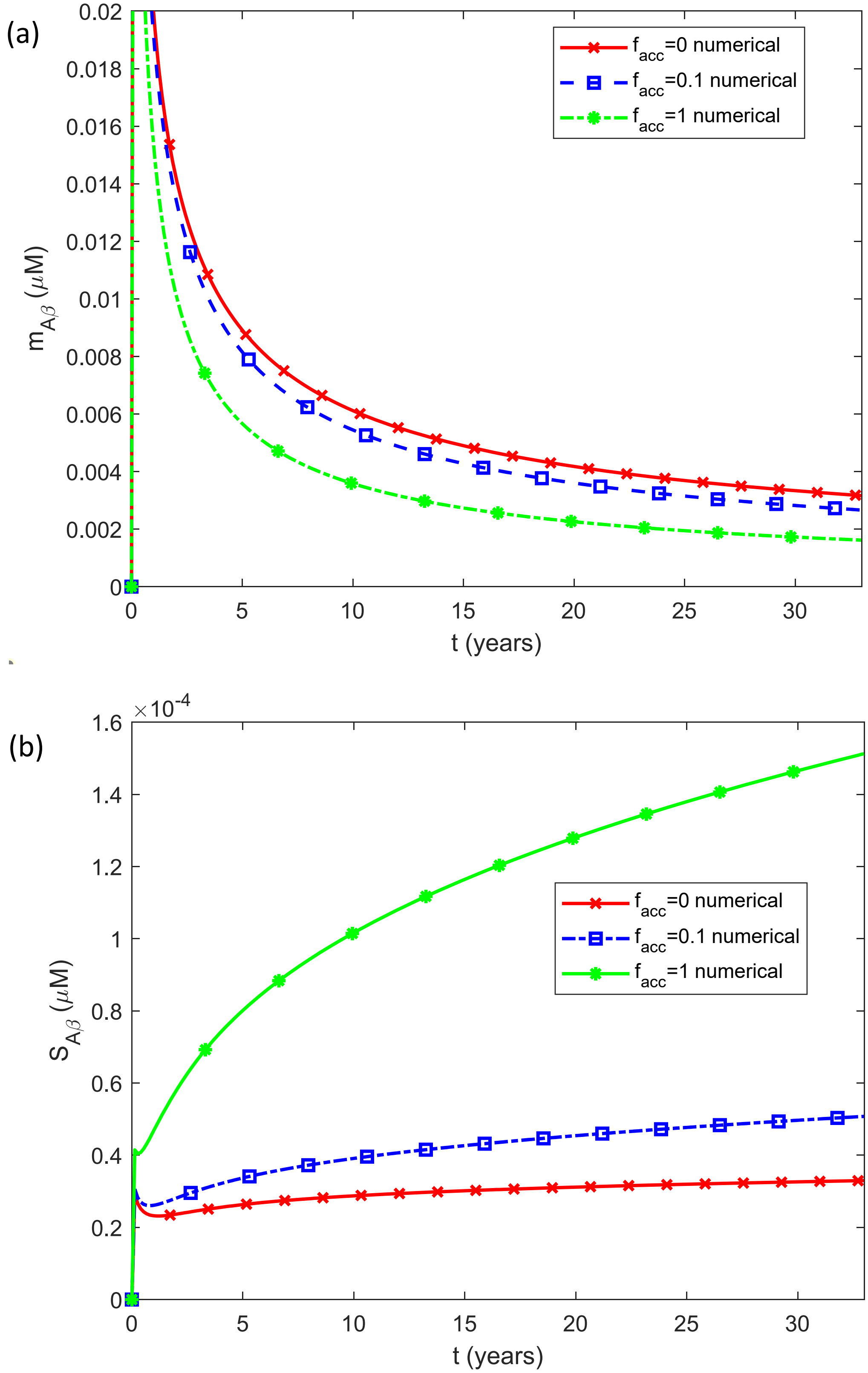

(a)
0.02
0.018
0.016
0.014
0.012
0.01
0.008
0.006
0.004
0.002
0
$m_{A\beta}$ ($\mu$M)
0
5
10
15
20
25
30
t (years)
$f_{acc}$=0 numerical
$f_{acc}$=0.1 numerical
$f_{acc}$=1 numerical
(b)
$\times10^{-4}$
1.6
1.4
1.2
1
0.8
0.6
0.4
0.2
0
$S_{A\beta}$ ($\mu$M)
0
5
10
15
20
25
30
t (years)
$f_{acc}$=0 numerical
$f_{acc}$=0.1 numerical
$f_{acc}$=1 numerical

Figure 2

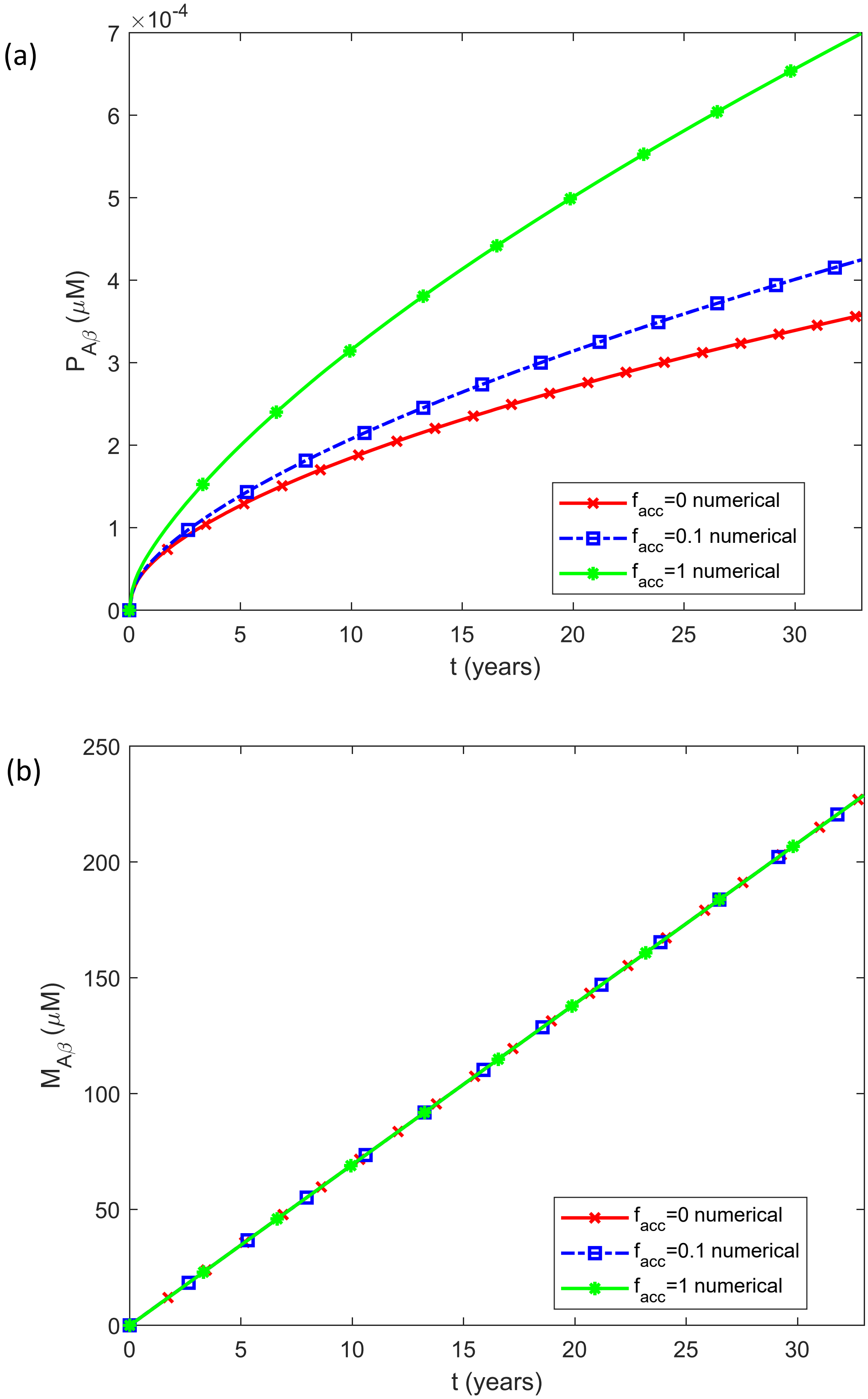

(a)
×10^-4
7
6
5
4
3
2
1
0
P_Aβ (μM)
0
5
10
15
20
25
30
t (years)
f_acc=0 numerical
f_acc=0.1 numerical
f_acc=1 numerical
(b)
250
200
150
100
50
0
M_Aβ (μM)
0
5
10
15
20
25
30
t (years)
f_acc=0 numerical
f_acc=0.1 numerical
f_acc=1 numerical

Figure 3

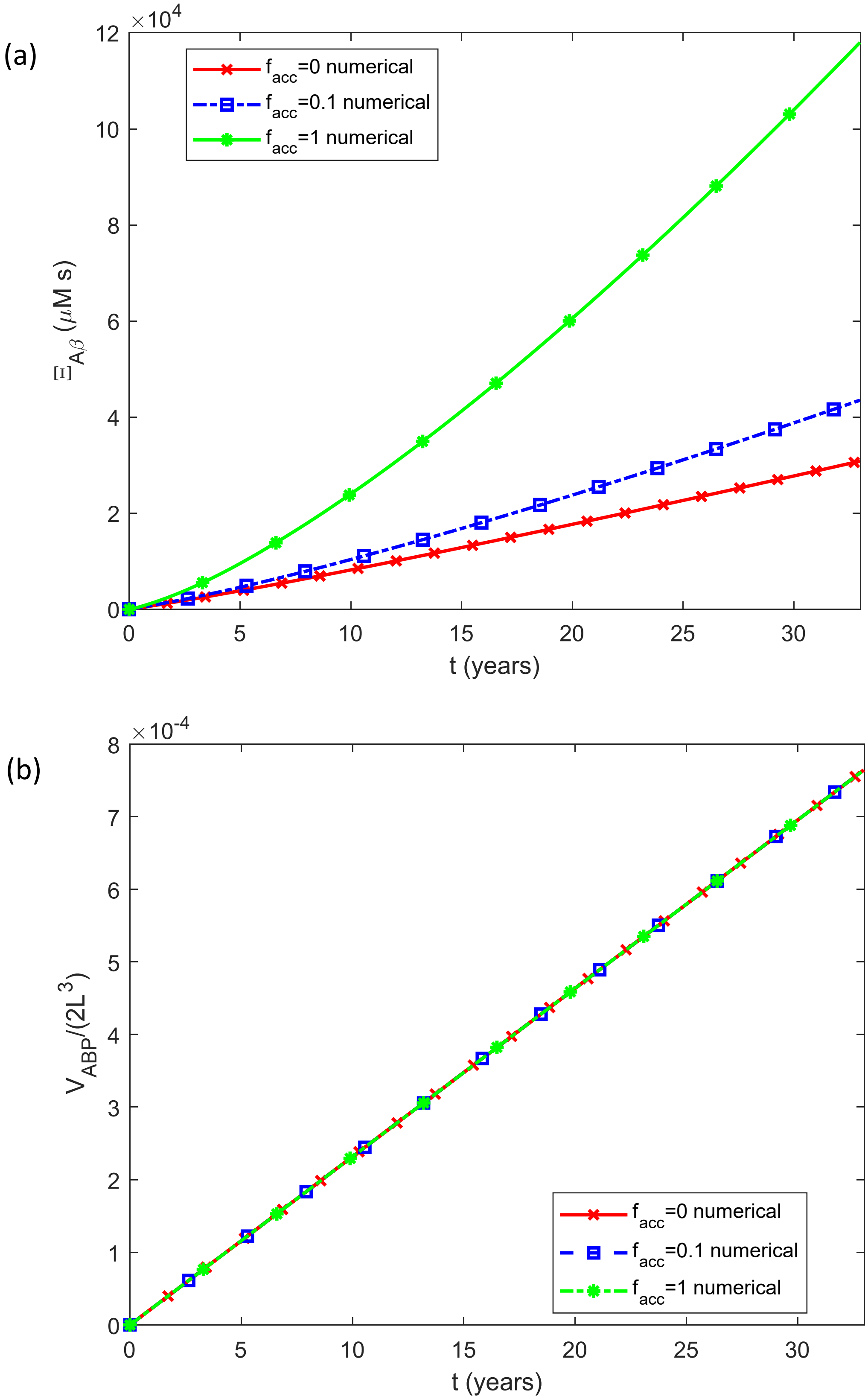

(a)
×10^4
12
10
8
6
4
2
0
$\Xi_{A\beta}$ ($\mu$M s)
$f_{acc}$=0 numerical
$f_{acc}$=0.1 numerical
$f_{acc}$=1 numerical
0
5
10
15
20
25
30
t (years)
(b)
×10^-4
8
7
6
5
4
3
2
1
0
$V_{ABP}/(2L^3)$
$f_{acc}$=0 numerical
$f_{acc}$=0.1 numerical
$f_{acc}$=1 numerical
0
5
10
15
20
25
30
t (years)

Figure 4

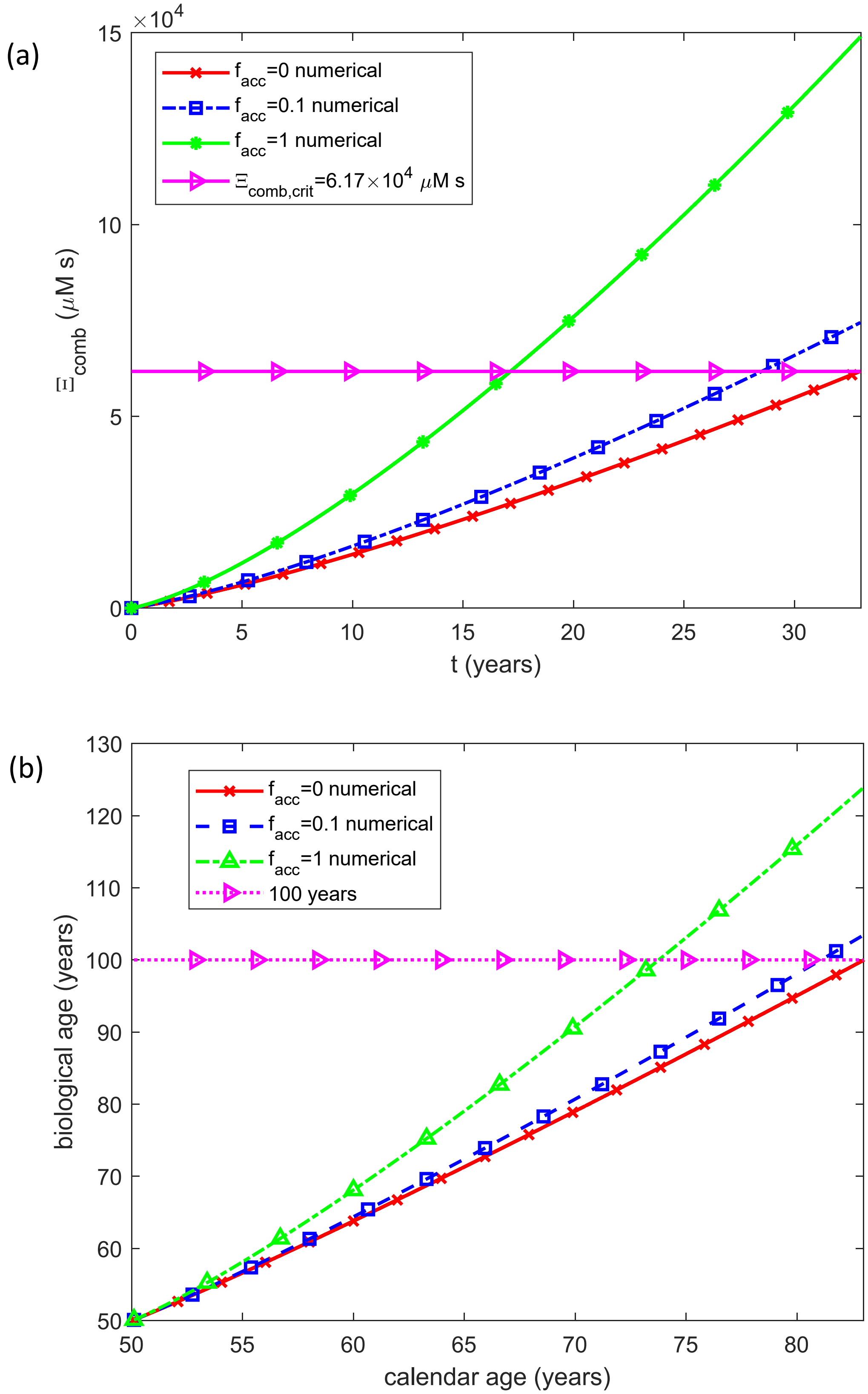

(a)
×10^4
15
10
5
0
$f_{acc}$=0 numerical
$f_{acc}$=0.1 numerical
$f_{acc}$=1 numerical
$\Xi_{comb,crit}$=6.17×10^4 μM s
$\Xi_{comb}$ (μM s)
0
5
10
15
20
25
30
t (years)
(b)
130
120
110
100
90
80
70
60
50
$f_{acc}$=0 numerical
$f_{acc}$=0.1 numerical
$f_{acc}$=1 numerical
100 years
biological age (years)
50
55
60
65
70
75
80
calendar age (years)

Figure 5

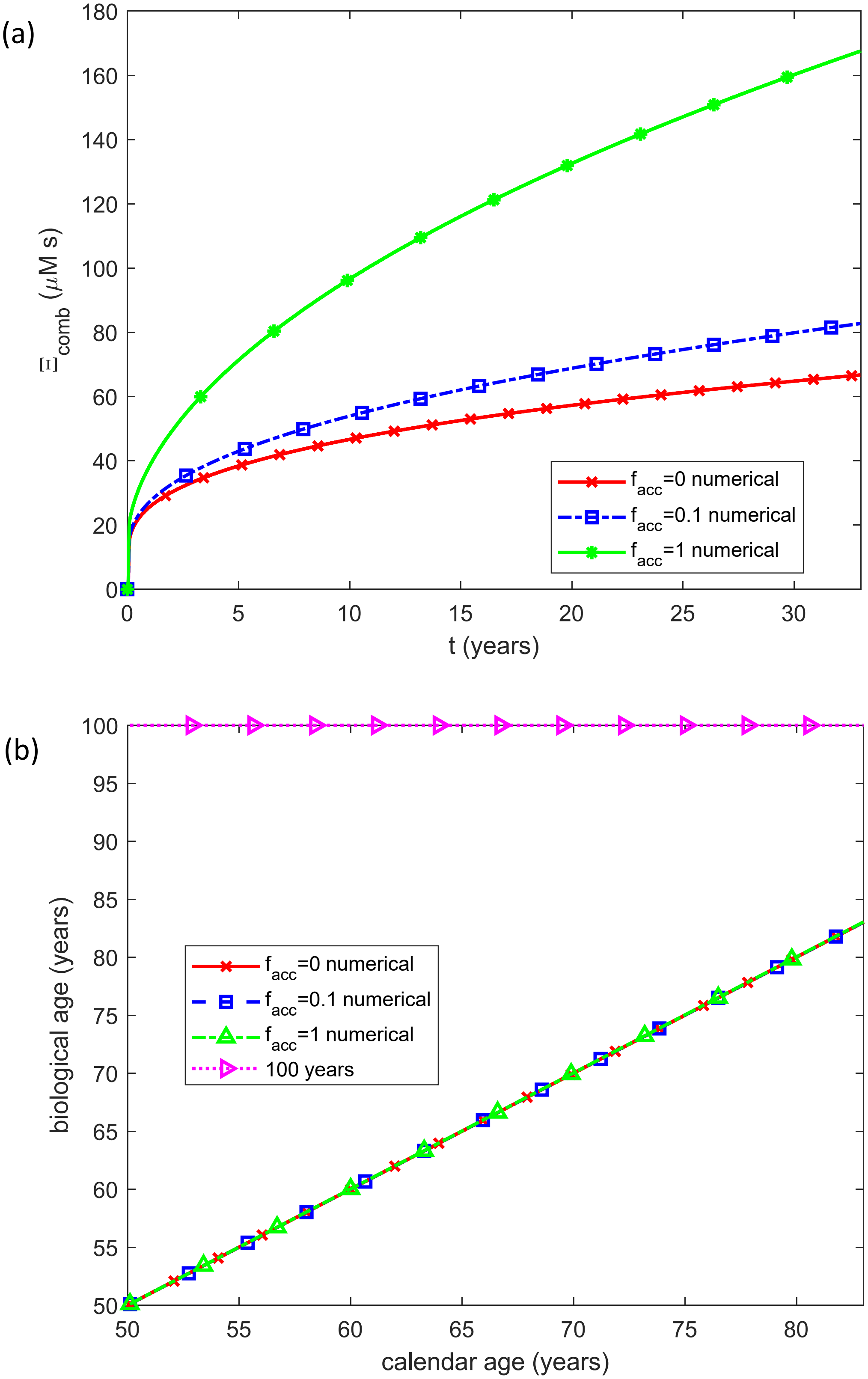

(a)
$\Xi_{comb}$ ($\mu$M s)
t (years)
$f_{acc}$=0 numerical
$f_{acc}$=0.1 numerical
$f_{acc}$=1 numerical
(b)
biological age (years)
calendar age (years)
$f_{acc}$=0 numerical
$f_{acc}$=0.1 numerical
$f_{acc}$=1 numerical
100 years

Figure 6

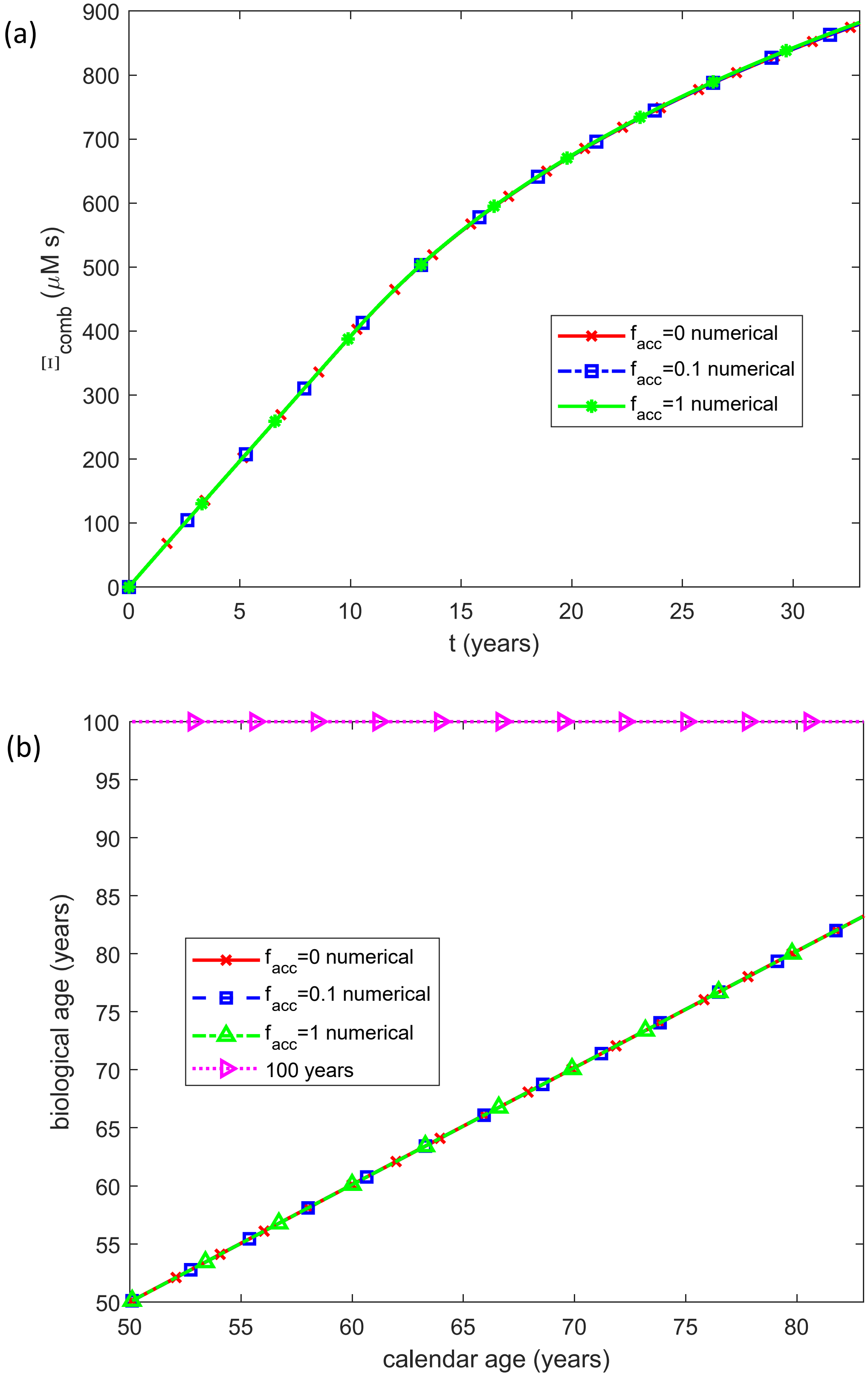
(a)
900
800
700
600
500
400
300
200
100
0
0
5
10
15
20
25
30
t (years)
Ξcomb (μM s)
facc=0 numerical
facc=0.1 numerical
facc=1 numerical
(b)
100
95
90
85
80
75
70
65
60
55
50
50
55
60
65
70
75
80
calendar age (years)
biological age (years)
facc=0 numerical
facc=0.1 numerical
facc=1 numerical
100 years

Figure 7

**A kinetic model predicts that tau-catalyzed Aβ nucleation can raise oligomer burden without increasing plaque volume**

Andrey V. Kuznetsov

Department of Mechanical and Aerospace Engineering, North Carolina State University,

Raleigh, NC 27695-7910, USA; email: avkuznet@ncsu.edu

## Supplemental Materials

### S1. Numerical Solution

The coupled system of ordinary differential equations, Eqs. (1)–(4), (10) and (12)–(14), subject to the initial conditions specified in Eqs. (5) and (15), was integrated numerically using the MATLAB solver ode15s (R2024a, MathWorks, Natick, MA, USA). ode15s is a variable-step, variable-order solver based on numerical differentiation formulas (NDFs) of orders 1–5 and is well suited to potentially stiff systems involving widely separated kinetic timescales. To ensure high numerical accuracy, both the relative and absolute error tolerances, RelTol and AbsTol, were set to $10^{-10}$. The accumulated exposures were evaluated on a uniform grid of $2\times10^4$ output points; convergence was verified by grid refinement, the dissociation scenario requiring at least $2\times10^3$ points because the oligomer concentration varies rapidly over the first months. At each integration step, the derived coupling variable $T_{tau}(t)$ and the intrinsic and homotypic Aβ nucleation rates, $R_{1,A\beta}$ and $R_{2,A\beta}$, were evaluated algebraically from the current solution. The tau-mediated heterotypic Aβ primary-nucleation rate was evaluated as $R_{tau\to A\beta} = \alpha_p T_{tau} m_{A\beta}^2$, using the fold-dependent coefficients $\alpha_p$ calibrated from the experimental data of (Mosconi et al., 2026).

The same comparison also bears on the choice of $k_{o2,A\beta}$. (Mosconi et al., 2026) report saturated multistep fits in which $k_2$ and $K_M$ appear only through their product over the monomer range studied, so that their two self-aggregation fits give $k_2$ values differing by a factor of 217 while the products $k_2$ $K_M$ agree to within 40 %. Those constants are therefore uninformative about $k_2$ at low monomer concentration, which is the quantity the present model requires. The informative constants are the unsaturated ones: $2.78\times10^{-9}$ $\mu M^{-2}s^{-1}$ from the secondary-nucleation fit of their Supplementary Table S1A, and $1.00\times10^{-8}$ $\mu M^{-2}s^{-1}$ from (Dear et al., 2020). These differ by a factor of 3.6 and form the two ends of the range given in Table 2. The closed-system predictions are insensitive to the choice within that range: the predicted half-time for 4 μM Aβ42 alone is 23.0 h with the lower value and 18.7 h with the upper, against 24.1 h for the saturated form, so the discrepancy with the measured value of about 5 h, and its attribution to the oligomer-to-fibril conversion step, holds in every case. The predicted lag-time acceleration at 1 µM PHF is likewise 2.22, 1.96 and 2.34 for the three choices, against a measured 2.50.

### S2. Closed-system consistency check of the heterotypic nucleation constant

The coefficient $\alpha_p$ is calibrated in Section 2.2 from the dependence of the fitted Aβ42 primary-nucleation rate constant on tau seed concentration, but the *in vivo* configuration of the main model cannot test it. In the corresponding tissue-scale model without the plaque-surface limitation of Eq. (11), a continuous monomer source leads to an Aβ fibril mass of approximately 229 µM, at which point homotypic secondary nucleation exceeds intrinsic primary nucleation by four orders of magnitude and no primary-nucleation channel, tau-catalyzed or otherwise, is resolvable in the predicted observables. The experiment reported by (Mosconi et al., 2026) supplied the tau-seeding measurements and several Aβ kinetic parameters used here, has the opposite character: a finite initial monomer load with no source, so that the fibril mass never exceeds that load and, during the lag phase, remains far below it. Reproducing that configuration allows the model to be compared against the measured kinetics rather than only fitted to them, and does so in the one regime where the comparison is possible.

The closed-system model uses Eqs. (10) and (12)–(14) with four changes. The volume-averaged source is set to zero; the monomer and oligomer half-lives are taken as infinite, the experiment lasting hours rather than decades; $T_{tau}$ is held constant, the tau seeds being preformed and not themselves growing, so that the tau compartment Eqs. (1)-(4) is not integrated; and homotypic secondary nucleation is written in the saturated multistep form actually fitted by (Mosconi et al., 2026), $R_{2,A\beta} = k_{o2,A\beta}\, m_{A\beta}^2\, M_{A\beta} / (1 + m_{A\beta}^2/K_{M,A\beta})$, with the constants of their Supplementary Table S1. The plaque-surface limitation of Eq. (11) is not applied, because *in vitro* the aggregate load is low and the fibrils are dispersed, so that catalytic surface scales with fibril mass. The experimental conditions represented are those of their Fig. 1: 4 µM monomeric Aβ42 in 20 mM phosphate buffer at pH 8 and 37 °C under quiescent conditions, with 0 to 1 µM preformed tau297-391 seeds of PHF or CTE fold, the seed concentration being expressed, as in that work, as a fibril mass concentration. The initial condition is $m_{A\beta}(0) = m_0$ with $S_{A\beta} = P_{A\beta} = M_{A\beta} = 0$, and the system conserves mass exactly, which was verified numerically to a relative error below $10^{-15}$.

The reason the tau-catalyzed channel is resolvable in this configuration is that during the lag phase the fibril mass is negligible, so the two primary-nucleation terms compete directly. The tau-catalyzed term equals the intrinsic one at $T_{tau} = k_{o1,A\beta}/\alpha_p$, which is 0.117 µM for PHF and 0.037 µM for CTE. Both thresholds lie at or below the lowest seed concentration tested experimentally, and the difference between them accounts for the observation of (Mosconi et al., 2026) that CTE seeds shortened the lag time significantly from 0.125 µM whereas PHF seeds did so only from 0.5 µM. In the corresponding in vivo model without the plaque-surface restriction of Eq. (11), the homotypic secondary-nucleation rate exceeds the intrinsic primary-nucleation rate within approximately ten days after the monomer source begins and remains larger thereafter.

Table S1 compares the predicted lag time with the values measured by (Mosconi et al., 2026), Fig. 1C, over the seed range for which $\alpha_p$ was calibrated. The lag time is evaluated, as in that work, as the intercept of the maximum-slope tangent to the fibril-mass curve.

*Table S1. Closed-system predictions for 4 µM Aβ42 with tau297-391 seeds, compared with the lag times measured by (Mosconi et al., 2026), Fig. 1C. Ratios are* $t_{lag}(0)/\ t_{lag}(T_{tau})$.

| tau seed (µM) | polymorph | $t_{1/2}$ (h) | $t_{lag}$ (h) | predicted ratio | measured ratio |
|---|---|---|---|---|---|
| 0 | – | 24.05 | 14.42 | 1.00 | 1.00 |
| 0.125 | PHF | 20.22 | 11.09 | 1.30 | 1.12 |
| 0.25 | PHF | 18.12 | 9.45 | 1.53 | – |
| 0.50 | PHF | 15.66 | 7.72 | 1.87 | 1.61 |
| 1.00 | PHF | 13.11 | 6.15 | 2.34 | 2.50 |
| 0.125 | CTE | 17.51 | 8.99 | 1.60 | 1.73 |
| 0.25 | CTE | 14.99 | 7.29 | 1.98 | – |
| 0.50 | CTE | 12.46 | 5.79 | 2.49 | 2.14 |
| 1.00 | CTE | 10.14 | 4.58 | 3.15 | 5.62 |

Within this calibration dataset, the agreement on the relative effect of tau is close. The predicted fold acceleration of the lag phase rises to 2.34 at 1 µM PHF against a measured 2.50, and the predicted values at 0.125 and 0.5 µM, 1.30 and 1.87, exceed the corresponding measured values of 1.12 and 1.61. The ordering of the two polymorphs is reproduced, CTE accelerating the reaction more than PHF at every seed concentration, as is the roughly threefold ratio of their catalytic coefficients. The model under-predicts the CTE effect at the highest seed concentration, 3.15 against a measured 5.62, which is consistent with the positive cooperativity that (Mosconi et al., 2026) report for both polymorphs and which the linear form of Eq. (9) does not represent. Within the range over which it was calibrated, $\alpha_p$ therefore reproduces the relative acceleration of Aβ42 nucleation within approximately 6–16 % for PHF across the tested seed concentrations, while the CTE predictions differ by up to a factor of two. Because the comparison uses measurements from the same study used for calibration, it is a within-dataset consistency check rather than independent validation.

The absolute timescale, by contrast, is not reproduced. The model predicts a half-time of 24.1 h for 4 μM Aβ42 alone, whereas the measured value is approximately 5 h and the reaction reaches its plateau within the 15 h of the experiment. The discrepancy is traced to the oligomer-to-fibril conversion step, which is taken from (Dear et al., 2020) and has no counterpart in the model of (Mosconi et al., 2026): raising $k_{conv,A\beta}$ from $1.83\times10^{-6}$ to $10^{-5}$ $\mu M^{-1}s^{-1}$ shortens the predicted half-time to 14.6 h, raising it further to $10^{-4}$ $\mu M^{-1}s^{-1}$ shortens it to 9.2 h, and removing the bottleneck entirely gives 7.7 h. This is a direct quantitative consequence of combining rate constants from two different global fits, and it is the kind of check that the *in vivo* configuration cannot supply. The predicted mean fibril length is affected in the same way: 3.5 μm against a measured value of approximately 500 nm for self-aggregated and 250 nm for seeded Aβ42, an overestimate consistent with the neglect of Aβ fibril fragmentation discussed in footnote j. The direction of the seeded change is nevertheless correct, the predicted mean fibril size falling by 12 % with PHF and 23 % with CTE at 1 μM seed.

(Dear et al., 2024), Eq. (2), provide an additional consistency relation that estimates the ratio of fibril-catalyzed to spontaneous oligomer dissociation. For Aβ42, $n_2 \simeq n_c \simeq 2$, so this ratio is independent of monomer concentration. Combined with the fitted total dissociation rate, the relation determines $k_{d1,A\beta}$ and $k_{d2,A\beta}$ individually rather than only their combination; the derivation and resulting values are given in notes b and d to Table 2. Both constants are set to zero in the baseline simulations, so this relation affects only the scenario in which oligomer dissociation is active.

The closed-system calculations used a MATLAB right-hand-side function and a separate driver script, distinct from those used for the *in vivo* model. Rate constants drawn from (Mosconi et al., 2026), Supplementary Table S1, are entered in the source units and converted in place, making the conversion from $M^{-1}h^{-1}$ to $\mu M^{-1}s^{-1}$ explicit and auditable. The driver integrates the system with ode15s at relative and absolute tolerances of $10^{-10}$ and $10^{-14}$, checks mass conservation at every output point, locates the half-time by interpolation and the lag time by the maximum-slope tangent, tabulates both against the measured values, and reports the two diagnostics described above.

**S3. Supplementary figures**

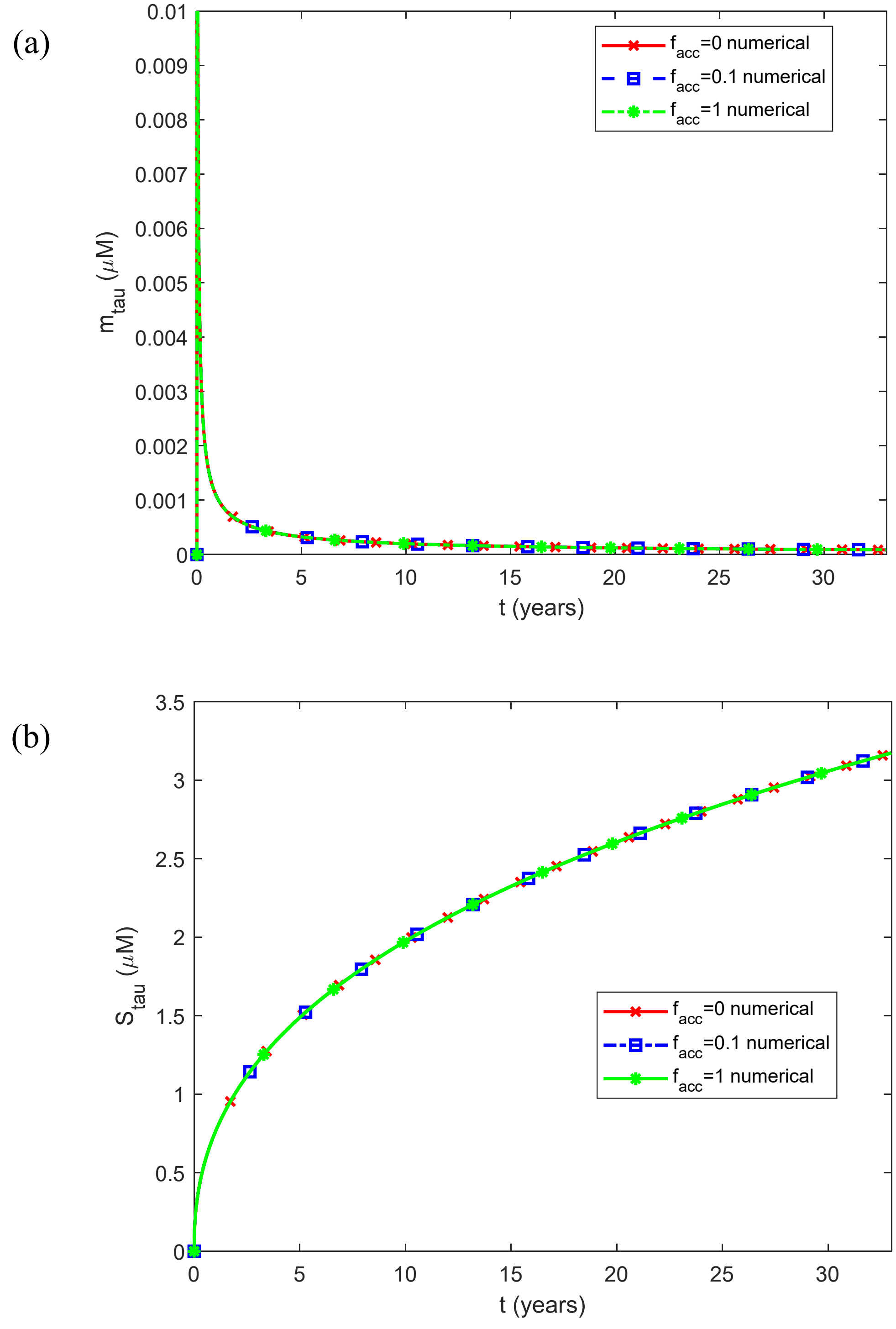

Fig. S1. (a) Molar concentration of free tau monomers as a function of time, $m_{tau}(t)$. (b) Molar concentration of free on-pathway tau oligomers as a function of time, $S_{tau}(t)$. Results are shown for zero

dissociation constants ( $k_{d1,tau}$ , $k_{d2,tau}$ , $k_{d1,A\beta}$ , $k_{d2,A\beta}$ ) and infinite half-lives ( $T_{1/2,m,tau}$ , $T_{1/2,S,tau}$ , $T_{1/2,m,A\beta}$ , $T_{1/2,S,A\beta}$ ).

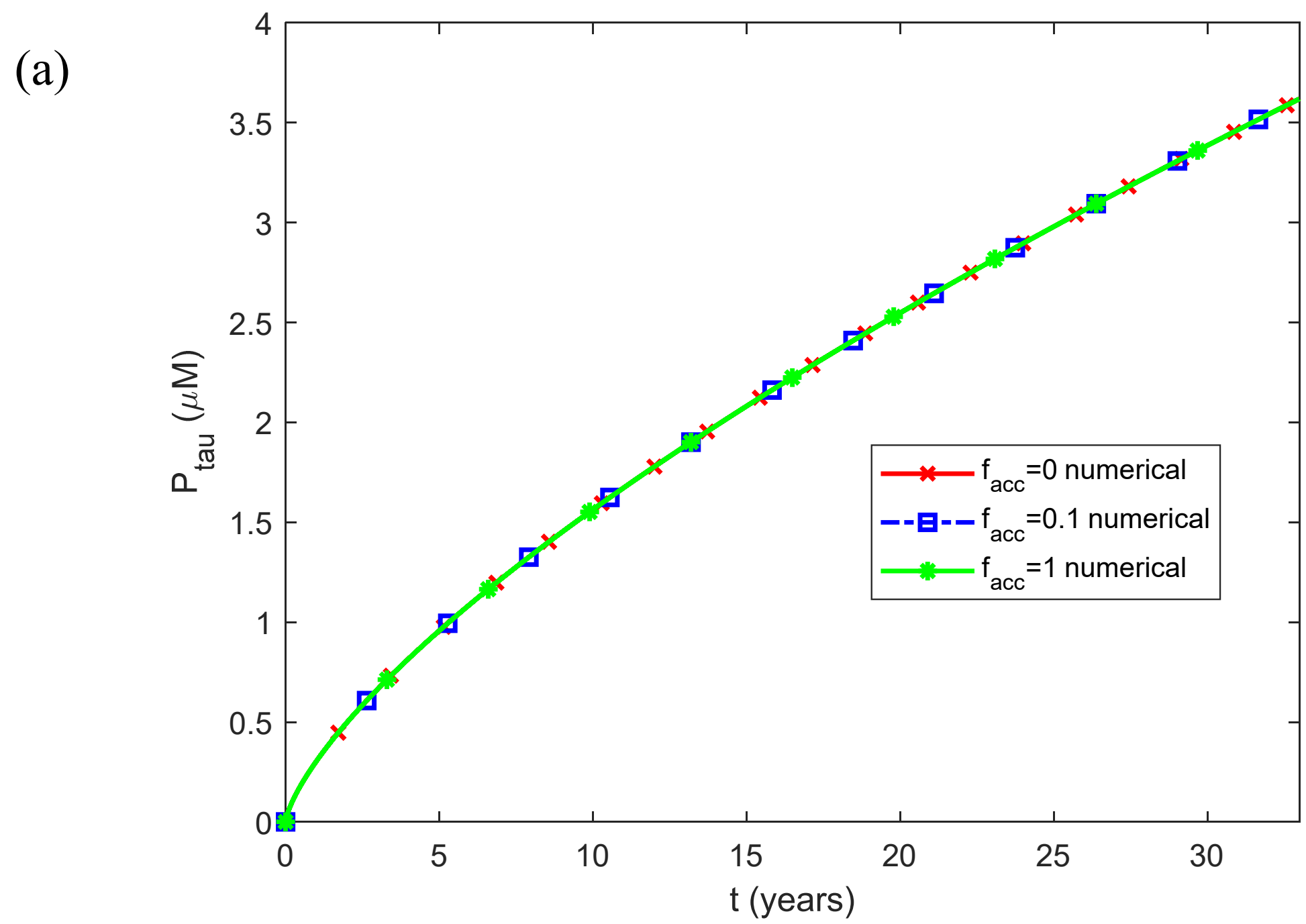

(b)

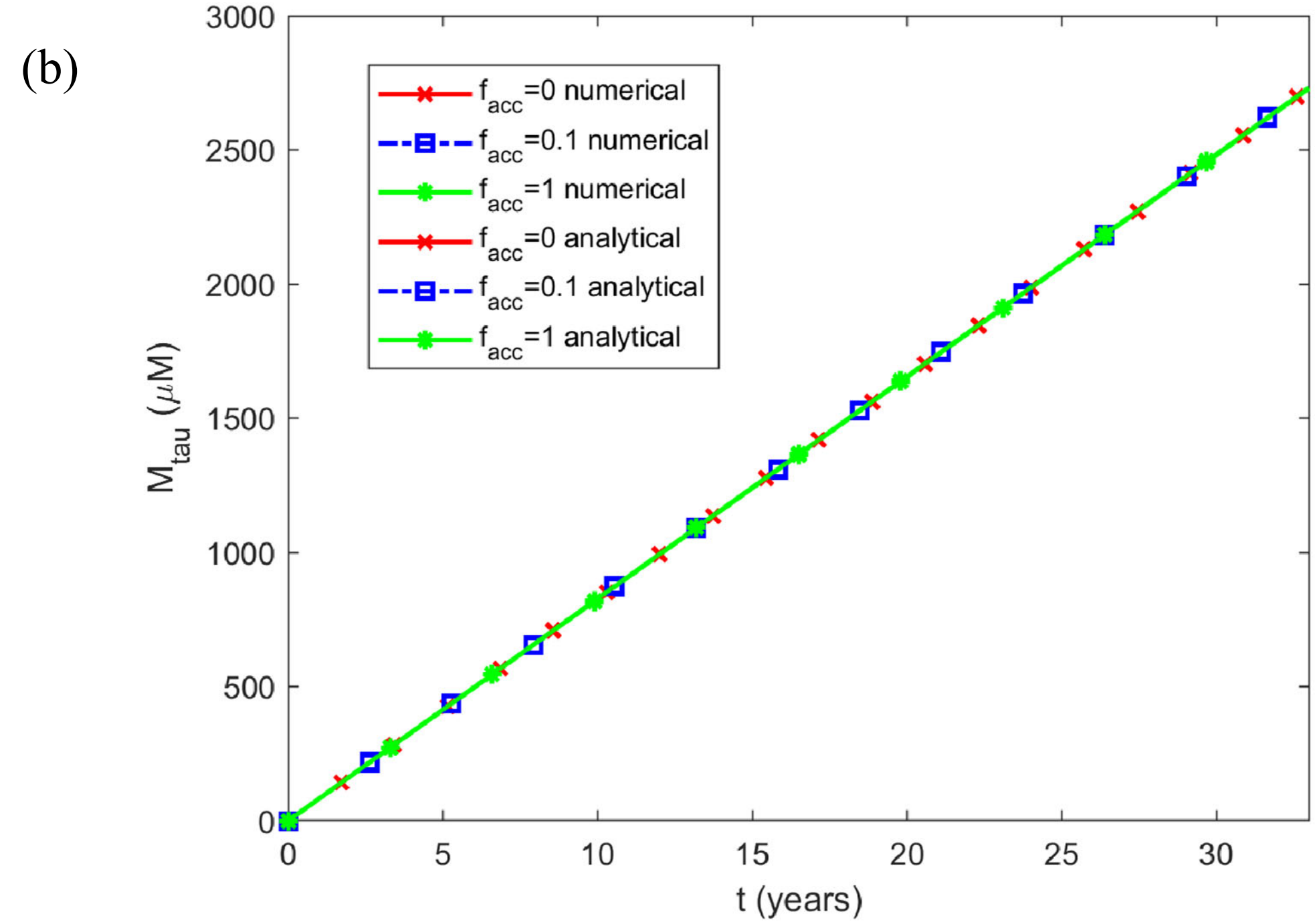

Fig. S2. (a) Molar concentration of tau fibrillar species of varying length as a function of time, $P_{tau}(t)$. (b) Molar concentration of tau monomers incorporated into fibrillar species of varying length as a function of time, $M_{tau}(t)$. Results are shown for zero dissociation constants ($k_{d1,tau}$, $k_{d2,tau}$, $k_{d1,A\beta}$, $k_{d2,A\beta}$) and infinite half-lives ($T_{1/2,m,tau}$, $T_{1/2,S,tau}$, $T_{1/2,m,A\beta}$, $T_{1/2,S,A\beta}$).

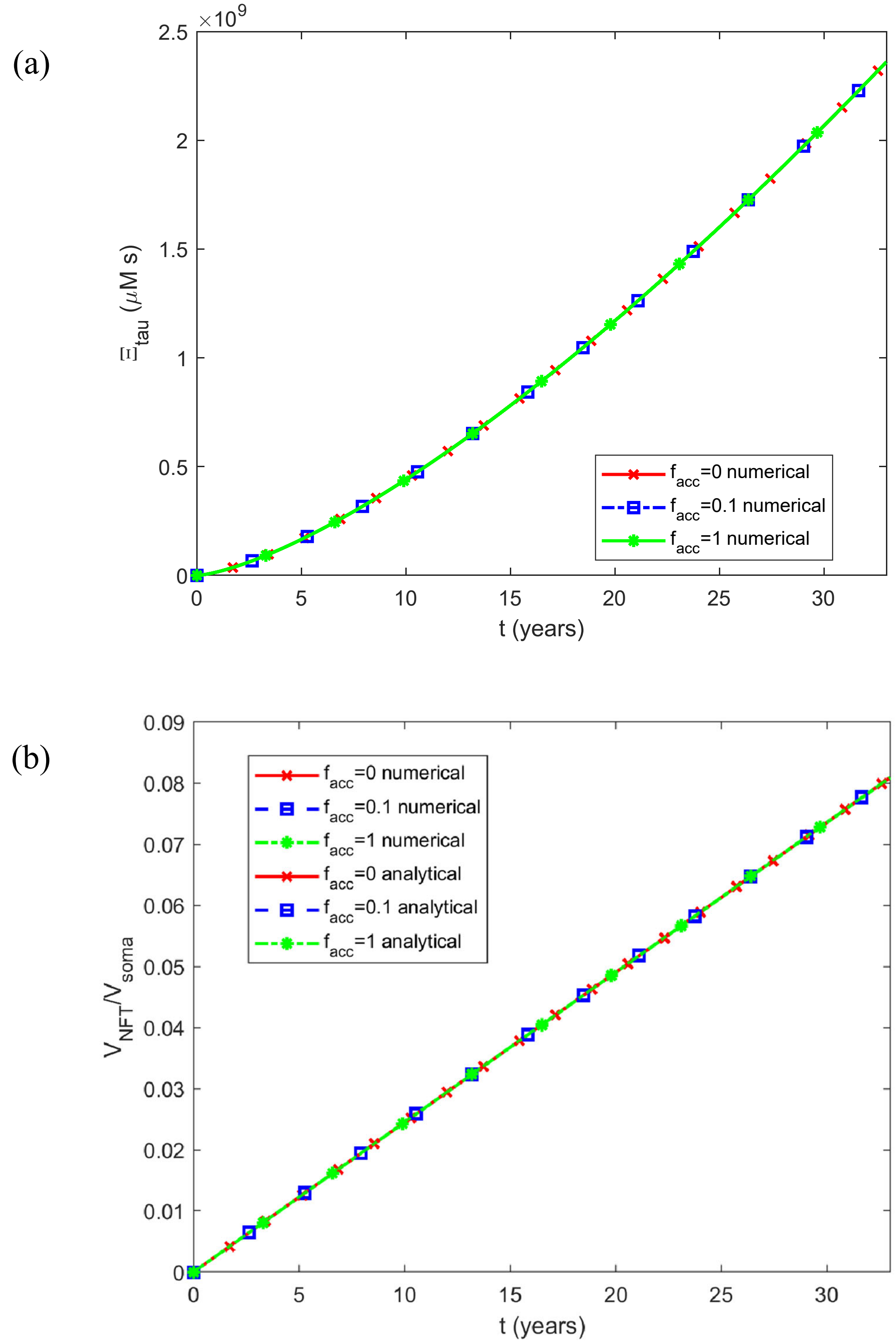

Fig. S3. (a) Accumulated neurotoxicity caused by tau oligomers as a function of time, $\Xi_{tau}(t)$. (b) Fraction of the CV occupied by neurofibrillary tangles as a function of time. Results are shown for zero

dissociation constants ( $k_{d1,tau}$ , $k_{d2,tau}$ , $k_{d1,A\beta}$ , $k_{d2,A\beta}$ ) and infinite half-lives ( $T_{1/2,m,tau}$ , $T_{1/2,S,tau}$ , $T_{1/2,m,A\beta}$ , $T_{1/2,S,A\beta}$ ).